\global\def\draftcontrol{0}

   \def\versionno{ Dp/Dq Phase Diagram -- draft   }

\catcode`\@=11

\expandafter\ifx\csname draftcontrol\endcsname\relax\global\def\draftcontrol{0}
\fi

{\count255=\time\divide\count255 by 60
\xdef\hourmin{\number\count255}
\multiply\count255 by-60\advance\count255 by\time
\xdef\hourmin{\hourmin:\ifnum\count255<10 0\fi\the\count255}}
\def\draftdate{\number\month/\number\day/\number\year\ \ \ \hourmin }

\newcommand\makepapertitle{\par
  \begingroup
    \renewcommand\thefootnote{\@fnsymbol\c@footnote}%
    \def\@makefnmark{\rlap{\@textsuperscript{\normalfont\@thefnmark}}}%
    \long\def\@makefntext##1{\parindent 1em\noindent
            \hb@xt@1.8em{%
                \hss\@textsuperscript{\normalfont\@thefnmark}}##1}%
     \newpage
     \global\@topnum\z@   
     \@makepapertitle
     \thispagestyle{empty}\@thanks
  \endgroup
  \setcounter{footnote}{0}%
  \global\let\thanks\relax
  \global\let\makepapertitle\relax
  \global\let\@makepapertitle\relax
  \global\let\@thanks\@empty
  \global\let\@author\@empty
  \global\let\@date\@empty
  \global\let\@title\@empty
  \global\let\title\relax
  \global\let\author\relax
  \global\let\date\relax
  \global\let\and\relax
  \def\version{\let\version\@version\@gobble}
}
\def\@makepapertitle{%
  \newpage
   \ifnum\draftcontrol=1 {}
   \version\versionno
   \vskip 3em%
   \else
   \hfill\hbox to 3cm {\parbox{4cm}{\@pubnum}\hss}%
   \vskip 3em%
   \fi
   \begin{center}%
   \let \footnote \thanks
     {\LARGE {\@title}}%
     \vskip 1.5em%
     {\normalsize
       \lineskip .5em%
       \begin{tabular}[t]{c}%
         \@author
       \end{tabular}\par}%
     \vskip 1.5em%
     {\@bstract}%
     \end{center}%
     \vskip 1.5em
     \@date%
   \par
}

\gdef\@pubnum{}
\def\pubnum#1{%
  \gdef\@pubnum{#1}}

\gdef\@bstract{}
\def\Abstract#1{%
  \gdef\@bstract{%
   \parbox{\textwidth-0pc}{%
   \centerline{\bf Abstract}\penalty1000%
\kern.2cm%
\noindent
\renewcommand\baselinestretch{1.0}%
{#1}}}
}

\def\ps@paper{\let\@mkboth\@gobbletwo%
     \ifnum\draftcontrol=1
    \def\@oddfoot{\hbox to \textwidth{\tiny \versionno \hfil\tiny\draftdate}%
    \hskip -\textwidth \hbox to \textwidth{\hfil\rm\thepage\hfil}}%
     \else\def\@oddfoot{\hbox to \textwidth{\hfil\rm\thepage\hfil}}
     \fi
     \let\@evenfoot\@oddfoot
}

\def\body{\clearpage
          \pagestyle{paper}
    }

\def\@version#1{\ifnum\draftcontrol=1
\typeout{}\typeout{#1}\typeout{}
\vskip3mm\centerline{\hbox{\fbox{\normalsize{\tt DRAFT -- #1 -- }
                   {\draftdate}}}}\vskip3mm
\fi}
\let\version\@version
\long\def\eqlabel#1{\ifnum\draftcontrol=1
                    \tag@false  
                    \tag*{(\theequation) \hbox to -0.2cm{\hspace{0cm}\small{#1}\hss}}
                    \refstepcounter{equation}
                    \edef\@currentlabel{\theequation}
                    \ltx@label{#1}          
                    \else
                    \label{#1}
                    \fi
                    }
\let\st@bibitem\@bibitem
\let\st@lbibitem\@lbibitem
\ifnum\draftcontrol=1
  \def\@bibitem#1{%
    \st@bibitem{#1}\a@@label{#1}\ignorespaces}
  \def\@lbibitem[#1]#2{%
    \st@lbibitem[#1]{#2}\a@@label{#2}\ignorespaces}
  \def\a@@label#1{%
    \gdef\a@lab{\smash{\normalfont\small#1}}
    \ifvmode
      \if@inlabel
        \global\setbox\@labels\hbox{%
          \llap{\a@lab\let\a@lab\relax
                \kern\@totalleftmargin\kern\marginparsep}%
          \box\@labels}%
      \fi
    \fi}
\fi

\documentclass[11pt,letterpaper]{article}

\usepackage{amsmath,amssymb,array,calc,epsfig}
\usepackage{cite}
\usepackage[
      colorlinks=true,
      linkcolor=red,  
      urlcolor=blue,    
      filecolor=red,     
      linkcolor=blue,
      citecolor = red,
      pdfstartview=FitV,
      pdftitle={},
        pdfauthor={Mukund Rangamani},
        pdfsubject={},
        pdfkeywords={},
        pdfpagemode=None,
        bookmarksopen=true    
      ]{hyperref}
      
\usepackage{graphicx}
\usepackage{subfigure}
\usepackage{epstopdf}
\usepackage{ccaption}

\ifnum\draftcontrol=1
\fi

\renewcommand\baselinestretch{1.25}
\renewcommand\section{\@startsection {section}{1}{\z@}%
                                   {-3.5ex \@plus -1ex \@minus -.2ex}%
                                   {2.3ex \@plus.2ex}%
                                   {\normalfont\large\bfseries}}
\renewcommand\subsection{\@startsection{subsection}{2}{\z@}%
                                   {-3.25ex\@plus -1ex \@minus -.2ex}%
                                   {1.5ex \@plus .2ex}%
                                   {\normalfont\normalsize\bfseries}}
\renewcommand\subsubsection{\@startsection{subsubsection}{3}{\z@}%
                                   {-3.25ex\@plus -1ex \@minus -.2ex}%
                                   {1.5ex \@plus .2ex}%
                                   {\normalfont\normalsize\it}}
\renewcommand\paragraph{\@startsection{paragraph}{4}{\z@}%
                                   {-3.25ex\@plus -1ex \@minus -.2ex}%
                                   {1.5ex \@plus .2ex}%
                                   {\normalfont\normalsize\bf}}

\numberwithin{equation}{section}

\def\revise#1       {\raisebox{-0em}{\rule{3pt}{1em}}%
                     \marginpar{\raisebox{.5em}{\vrule width3pt\
                     \vrule width0pt height 0pt depth0.5em
                     \hbox to 0cm{\hspace{0cm}{%
                     \parbox[t]{4em}{\raggedright\footnotesize{#1}}}\hss}}}}

\def\sqr#1#2{{\vcenter{\vbox{\hrule height.#2pt
 \hbox{\vrule width.#2pt height#1pt \kern#1pt
 \vrule width.#2pt}\hrule height.#2pt}}}}

\def\aa1{\phi}
\def\cc1{\psi}

\catcode`\@=12

\begin{document}


\title{\bf Universality of Holographic Phase Transitions and Holographic Quantum Liquids}

\pubnum{%
arXiv:0911.xxxx}
\date{October 2009}

\author{
\scshape Paolo Benincasa\\[0.4cm]
\ttfamily Center for Particle Theory \& Department of Mathematical Sciences\\
\ttfamily Science Laboratories, South Road, Durham DH1 3LE, United Kingdom\\[0.2cm]
\small \ttfamily paolo.benincasa@durham.ac.uk
}

\Abstract{We explore the phase structure for defect theories in full generality using the gauge/gravity
correspondence. On the gravity side, the systems are constructed by introducing $M$ (probe) 
D$(p+4-2k)$-branes in a background generated by $N$ D$p$-branes to obtain a codimension-$k$ intersection. 
The dual gauge theory is a $U(N)$ Supersymmetric Yang-Mills theory on a $(1+p-k)$-dimensional defect with
both adjoint and fundamental degrees of freedom. We focus on the phase structure in the chemical potential
versus temperature $(\mu,\,T)$ plane. We observe the existence of two universality classes for holographic 
gauge theories, which are identified by the order of the phase transition in the interior of the 
$(\mu,\,T)$-plane. Specifically, all the {\it sensible} systems with no defect show a third order phase 
transition. Gauge theories on a defect with $(p-1)$-spatial directions are instead characterised by a second 
order phase transition. One can therefore state that the order of the phase transition in the interior of the 
$(\mu,\,T)$-plane is intimately related to the codimensionality of the defect. We also discuss the massless 
hypermultiplet at low temperature, where a thermodynamical instability seems to appear for $p<3$. Finally, we 
comment on such an instability.}

\makepapertitle

\body

\version\versionno

\section{Introduction}\label{Intro}

Gauge/gravity correspondence \cite{Maldacena:1997re, Gubser:1998bc, Witten:1998qj, Aharony:1999ti}
provides a powerful tool to investigate the dynamics of strongly coupled
gauge theories. The original formulation \cite{Maldacena:1997re} conjectures the equivalence between
supergravity on the ``near-horizon'' geometry generated by a stack of $N$ coincident D$3$-branes 
($AdS_{5}\times S^5$) and the gauge theory 
(four-dimensional $\mathcal{N}=4$ $SU(N)$ Supersymmetric Yang-Mills) living on the 
boundary of $AdS_{5}$, which describes the brane modes decoupled from the bulk. 
It can be straightforwardly extended to any asymptotically $AdS\times\mathcal{M}$ geometry, 
$\mathcal{M}$ being a compact manifold. This conjectured equivalence
is made precise by identifying the string partition function with the generating function for the gauge
theory correlators, with the boundary value of the bulk modes acting as source of the correspondent
gauge theory operator \cite{Witten:1998qj}.

It can be extended to the case of arbitrary D$p$-branes ($p\neq3$), for which the 
world-volume gauge theory is again equivalent to the supergravity on the near-horizon
background generated by the D$p$-branes \cite{Itzhaki:1998dd}. The dual gauge theory is a $(p+1)$-dimensional
$U(N)$ supersymmetric Yang-Mills theory. Contrarily to the case of the D$3$-branes, it has a dimensionful 
coupling constant and the effective coupling depends on the energy scale: the gauge theory is no longer 
conformal. More specifically, there exists a frame \cite{Duff:1994fg} in which the near-horizon 
geometry induced by the D$p$-branes is conformally $AdS_{p+2}\times S^{8-p}$ 
\cite{Boonstra:1997dy, Boonstra:1998yu, Boonstra:1998mp}. In this frame, the existence of a generalized
conformal symmetry \cite{Jevicki:1998ub} becomes manifest and the radial direction 
(transverse to the boundary) acquires the meaning of energy scale of the dual gauge theory 
\cite{Boonstra:1998mp, Skenderis:1998dq}, as in the original $AdS/CFT$-correspondence. 
Moreover, the holographic RG flow turns out to be trivial and the theory flows just because of the 
dimensionality of the coupling constant. In the case of the D$4$-branes, the theory flows to a 
$6$-dimensional fixed point at strong coupling: the world-volume theory of D$4$-branes flows to the 
world-volume theory of M$5$-branes.

Gauge/gravity correspondence can be further generalized by inserting extra degrees of freedom in the theory.
More precisely, one can add a finite number of branes and consider the probe approximation, so that the
backreaction on the background geometry can be neglected. Inserting probe branes introduces a fundamental 
hypermultiplet in the gauge theory, partially or completely breaking the original supersymmetries
\cite{Karch:2002sh}.

Here we are mainly interested in the phase structure of the BPS brane intersections at finite temperature and
finite chemical potential. The phase diagram temperature versus chemical potential has been studied in details 
especially in relation to the D$3$/D$7$ system, where the probe D$7$-branes are parallel to the background 
D$3$-branes \cite{Babington:2003vm, Mateos:2006nu, Albash:2006ew, Kobayashi:2006sb, Mateos:2007vn, 
Ghoroku:2007re, Mateos:2007vc, Faulkner:2008hm}. A similar analysis has been carried out for the D$4$/D$6$
system with the direction of the background D$4$-branes which is not parallel to the probe D$6$-branes
compactified to a circle so that the system has effectively codimension $0$ \cite{Matsuura:2007zx}.
Considering a system at finite temperature and finite 
(``baryonic'') chemical potential means considering a black brane background solution and a non-trivial 
(``electric'') profile for the world-volume gauge field on the probe branes: the temperature of the system is 
given by the Hawking temperature of the black brane background and the chemical potential is provided by the 
boundary value of the time component of the world-volume gauge potential. 

\begin{figure}[htbp]
 \centering%
 \subfigure[\protect\url{Minkowski-Embedding}\label{fig:Mink}]%
   {\scalebox{.6}{\includegraphics{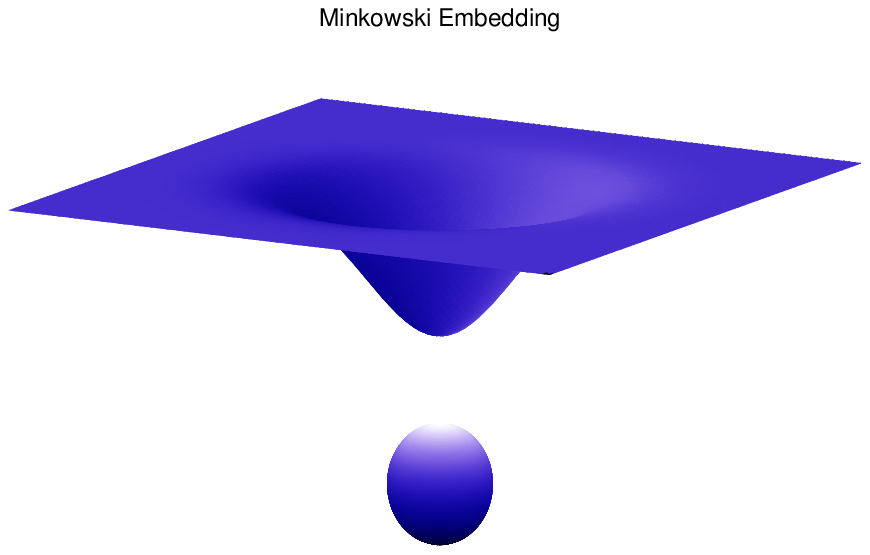}}}
 \subfigure[\protect\url{Critical-Embedding}\label{fig:Crit}]%
   {\scalebox{.6}{\includegraphics{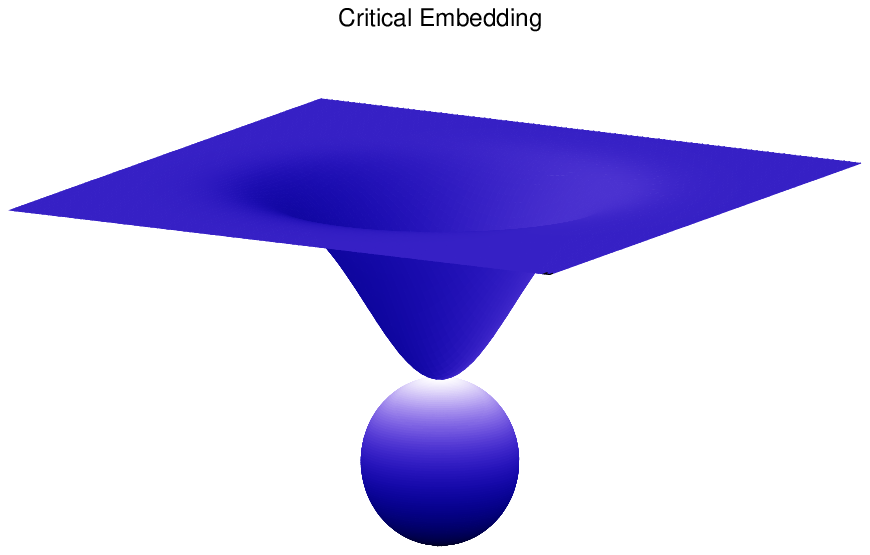}}}
 \subfigure[\protect\url{Black-Hole-Embedding}\label{fig:BH}]%
   {\scalebox{.6}{\includegraphics{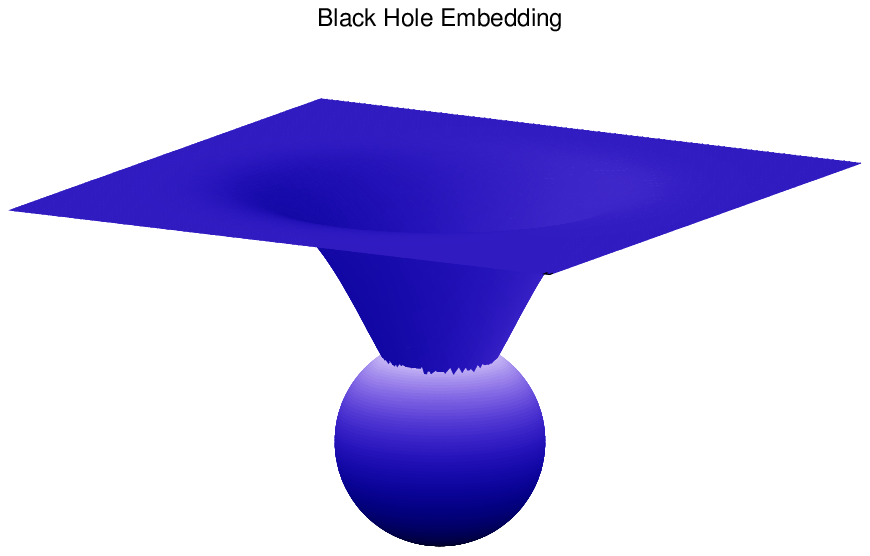}}}
 \caption{First order phase transition. At low enough temperature ($T\,<\,T_{\mbox{\tiny d}}$) the probe 
          branes lie outside the black hole (Minkowski embedding). Increasing the temperature, the black
          hole increases in size and the probe branes starts to bend towards it more and more
          until they touch the black hole at $T=T_{\mbox{\tiny d}}$ (Critical embedding). For
          $T>T_{\mbox{\tiny d}}$ part of the brane falls inside the black hole (Black hole embedding).
          The transition from Minkowski embedding to black-hole one is of first order.\label{fig:First}}
\end{figure}

Along the temperature axis, the system undergoes a first order phase transition (Figure \ref{fig:First}), 
which can be seen as a meson dissociation transition \cite{Mateos:2006nu, Hoyos:2006gb, Mateos:2007vn}.
From the branes perspective, at sufficiently small temperatures, the probe branes lie outside the black-hole 
background (Minkowski embedding - Figure \ref{fig:Mink}). Increasing the temperature, the size of the black 
hole increases and, consequently, the probe branes start to feel stronger and stronger the black hole 
attraction, bending more and more towards the black hole until they touch the black
hole horizon at one point (critical embedding) at temperature $T\,=\,T_{\mbox{\tiny $d$}}$ 
(Figure \ref{fig:Crit}). Continuing increasing the temperature part, of the probe branes fall into the black 
hole horizon (black hole embedding - Figure \ref{fig:BH}).

\begin{figure}[htbp]
 \centering%
 \subfigure[\protect\url{Minkowski-Embedding}\label{fig:Mink2}]%
   {\scalebox{.6}{\includegraphics{Mink_Emb_01.eps}}}
 \subfigure[\protect\url{Instanton-Condensation}\label{fig:Instant}]%
   {\scalebox{.6}{\includegraphics{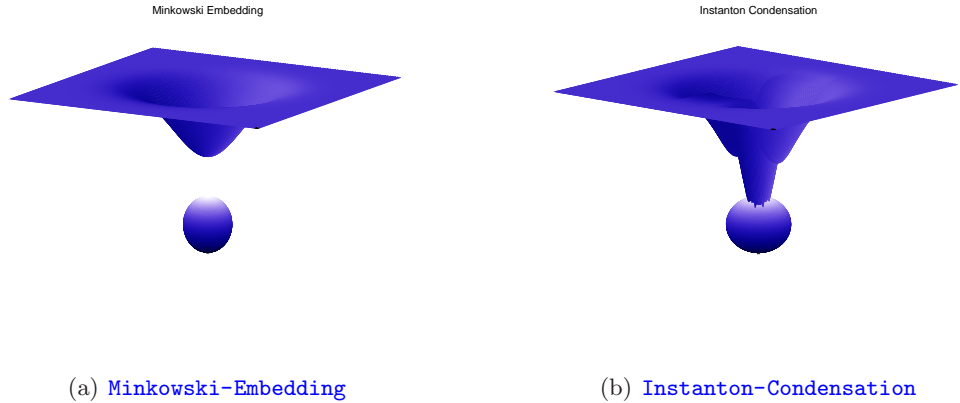}}}
 \caption{Worldsheet instanton effects. In the region $\mu\,>\,m$ the instanton corrections to
          the DBI-action becomes relevant given their dependence on 
          $\,\sim\,e^{-|n|\beta\left(m(T)-\frac{|n|}{n}\mu\right)}$. Their contribution becomes dominant
          so that they condense following a neck until the black hole, sending the system to
          the black hole phase.}\label{fig:Inst}
\end{figure}

Along the chemical potential axis, the system instead undergoes a second order phase transition 
\cite{Karch:2007br}. In the interior of the $\left(\mu,\,T\right)$-plane there has been identified
a transition curve $\mu\:=\:m\left(T\right)$: below this curve the system is in a Minkowski phase
in which the ``quark'' density is zero, while above it the system is in a black-hole phase with
non-vanishing ``quark'' density. This transition curve was numerically found to be first order
\cite{Kobayashi:2006sb}. However, recently an analytic computation by Faulkner and Liu 
\cite{Faulkner:2008hm} showed that this transition is actually of third order. In this picture, in the
region $\mu\,<\,m\left(T\right)$ of the phase diagram (Minkowski embedding), the DBI-action of the 
D$7$-branes provides the dominant contribution to description of the brane embedding. For 
$\mu\,>\,m\left(T\right)$  string worldsheet instantons contribute as well \cite{Faulkner:2008qk}, creating 
an instability: the instantons condense and create a neck between the probe D$7$-branes and the black hole so 
that at $\mu\,=\,m\left(T\right)$ the Minkowski embedding goes over to a black-hole embedding 
(Figure \ref{fig:Inst}).

A natural question to ask is how universal are these phase transitions and which type of phase transitions 
these systems allow to investigate. The intersecting brane constructions provide an arena to construct model 
which may points towards a deeper theoretical understanding of QCD-like 
\cite{Wang:2003yc, Kruczenski:2003uq, Sakai:2004cn, Sakai:2005yt, Burrington:2009fm} and 
condensed-matter-like \cite{Davis:2008nv, Ammon:2008fc, Myers:2008me, Fujita:2009kw, Ammon:2009fe, 
Peeters:2009sr, Wapler:2009tr, Hung:2009qk} features.

In this paper, we investigate the phase structure in the plane chemical potential versus temperature for
BPS intersecting brane systems, which are constructed by introducing a stack of $M$ probe D$(p+4-2k)$-branes
in a background generated by $N$ D$p$-branes ($M\ll N$). The parameter $k$ ($k\,=\,0,\,1,\,2$) indicates the 
codimensionality of the intersection, {\it i.e.} the number of the spatial directions of the background branes
along which the probe branes do not extend. The dual gauge theory is therefore a $(p+1)$-dimensional $U(N)$ 
Supersymmetric Yang-Mills theory with a $(p+1-k)$-dimensional defect on which the fundamental degrees of 
freedom propagate.

We start with considering the massive fundamental hypermultiplet, which means describing the embedding of
the probe branes through a coordinate of the transverse space. In order to introduce the chemical potential,
we turn on an electric ansatz for the world-volume gauge field. 

At zero temperature one can identify a brane/anti-brane phase and a ``black-hole'' crossing phase. In the 
first case, the probe brane can extend from the boundary to a minimum distance at which the branes turn and go
back hitting the boundary again, forming therefore a brane/anti-brane configuration. In the second case, the 
probe branes can extend down to the location of the background branes. The latter phase is thermodynamically 
favoured, the grand-potential being negative. The phase transition between these two configurations is of 
second order as in \cite{Karch:2007br}. 
The presence of a second order phase transition along the chemical potential 
axis is therefore a universal property of the D$p$/D$(p+4-2k)$ system. The previous analysis applies to 
systems with $k\,=\,0,\,1$. The case $k\,=\,2$ is very different and subtle: turning on the embedding mode in the 
transverse space does no longer correspond to a mass deformation, but rather it provides a 
vacuum-expectation-value for the dual operator. However, suitably defining the ``physical'' chemical potential as well
as the ``physical'' mass, one can show that also gauge theories dual to these brane constructions fall in this
universality class.

At finite temperature, the systems show a transition line in the chemical potential versus temperature plane.
We analyse such a phase transition analytically following the approach proposed in \cite{Faulkner:2008hm}.
The equations of motion for the embedding function and the gauge potential are too involved to be analytically
solved. The idea in \cite{Faulkner:2008hm} is to perturbatively solve them considering the ``quark''-density
(suitably rescaled to be a dimensionless parameter) as parameter expansion. Such a perturbative analysis 
allows to obtain an analytical expression for the chemical potential which in turn allows to analytically 
study the order of the phase transition. 
In the case of systems with no defect, the chemical potential acquires the following expression up
to first order in the quark density $c_{\mbox{\tiny $f$}}$.
\begin{equation}\eqlabel{cpfin}
 \mu\:=\:m\left(T\right)+\mathfrak{s}_{\mbox{\tiny $1$}}\left(T\right)c_{\mbox{\tiny $f$}}
         -\mathfrak{s}_{\mbox{\tiny $2$}}\left(T\right)c_{\mbox{\tiny $f$}}\log{c_{\mbox{\tiny $f$}}}
         +\mathcal{O}\left(c_{\mbox{\tiny $f$}}^2\right),
\end{equation}
{\it i.e.} the structure found in \cite{Faulkner:2008hm} for the D$3$/D$7$ system generalises to more
general D$p$/D$(p+4)$ systems.

In the case of codimension-$1$ defect systems, there is no logarithmic term at first order. 
This is actually crucial. The presence of the logarithmic term in \eqref{cpfin} implies that the phase transition is a of 
third order. In the case no logarithmic term is present at first order in $c_{\mbox{\tiny $f$}}$, 
the system show a second order phase transition. It is thus possible to state
that the order of the transition line in the chemical potential versus temperature plane is strictly tied
to the codimensionality of the system under examination: systems with a codimension-$0$ defect ({\it i.e.} no
defect) show a third order phase transition, while systems with a codimension-$1$ defect are characterised
by a second order phase transition. In the latter case, the whole transition line is of second order.
For $k=0$, the 
function $\mathfrak{s}_{\mbox{\tiny $2$}}$ has a point at $T\,=\,T_{\mbox{\tiny c}}\,<\,T_{\mbox{\tiny d}}$
where it vanishes: this point represents a tricritical point at which the phase transition becomes of 
second order.

We also consider the case of massless hypermultiplets in the fundamental representation. As mentioned before,
such description is provided by fixing the probe branes to wrap the maximal $S^{3-k}\,\subset\,S^{8-p}$ with
the probe branes wrapping in a $(p+2-k)$-dimensional subspace of the $(p+2)$-dimensional non-compact manifold.

For defect theories, this embedding of the probe branes is described by the scalar mode 
$x^{\mbox{\tiny $p$}}\,\equiv\,z\left(\rho\right)$, with $x^{\mbox{\tiny $p$}}$ being the direction of the
background branes along which the probe branes do not extend. When such mode has a non trivial profile, 
the supersymmetries are broken since it provides a vacuum-expectation-value to its dual operator. The 
supersymmetries are instead preserved if the scalar mode $z\left(\rho\right)$ has a trivial profile 
({\it i.e.} no scalar mode is turned on). For codimension-$0$ ($k=0$) systems, the supersymmetric embeddings 
are the only possible. In
\cite{Karch:2008fa} massless $\mathcal{N}=2$ hypermultiplets in $\mathcal{N}=4$ Supersymmetric
Yang Mills were studied via probe $D(7-2k)$-branes wrapping an 
$AdS_{\mbox{\tiny $5-k$}}\times S^{\mbox{\tiny $3-k$}}$ space. In this framework, the specific heat and the
zero-sound mode for holographic quantum fluids have been discussed, {\it i.e. } translationally invariant 
systems at low temperature and finite chemical potential. The zero sound mode turns out to persist to
all the value of the hypermultiplet mass for such systems \cite{Kulaxizi:2008kv} and was extensively discussed
for D$4$/D$8$/$\bar{\mbox{D}}8$ systems \cite{Kulaxizi:2008jx}. In \cite{Karch:2009eb} the analysis of 
\cite{Karch:2008fa} was extended to D$p$/D$q$ systems, where both massless and massive deformations were 
analysed. In real world, there are two main classes of
quantum liquids in $(1+d)$-dimensions, with $d>1$: the Bose and Fermi fluids. In the first case, the low 
energy elementary excitations are given by superfluids phonons with linear dispersion relation and specific
heat $c_{\mbox{\tiny $v$}}\,\sim\,T^{d}$. Fermi liquids have both bosonic and fermionic 
quasi-particle excitations, with the latter dominating at low temperature and fixing the specific heat to scale
directly proportionally to the temperature: $c_{\mbox{\tiny $v$}}\,\sim\,T$ ($\forall\:d\,>\,1$). This
Fermi-liquid type of behaviour has been observed in holographic gravity duals constructed with background
D$4$-branes \cite{Kulaxizi:2008jx, Karch:2009eb}. The $(1+1)$-dimensional case
is indeed peculiar and can never be described through Bose or Fermi liquids. First, in $(1+1)$-dimensions just
collective motion is possible. The reason is straightforward to understand. The excitations have one-direction 
only in which they can propagate and therefore they would necessarily scatter with other excitations put them 
in motion. Such systems are thought to be described by the Tomonaga-Luttinger liquids 
\cite{Tomonaga:1950aa, Luttinger:1963aa, Mattis:1965aa, Haldane:1981aa, Schulz:1998aa, Giamarchi:2009aa}. 
They do not have quasi-particle peaks and are 
characterised by two elementary excitations: plasmons, {\it i.e. } sound waves as response of charge density 
to external perturbations with speed dependent on the interactions, and spin density waves which propagates
independently of the plasmons. A first attempt to extensively discuss a holographic realisation of 
$(1+1)$-dimensional quantum liquids has been made using a D$3$/D$3$-system \cite{Hung:2009qk}. At low
temperature, the specific heat in the dual field theory scales quadratically in the temperature, while
for Tomonaga-Luttinger liquid the specific heat scales linearly if the system is not interacting and scales
as $\sim\,T^{\alpha(K)}$ for interacting systems (the power $\alpha(K)$ depends on the Luttinger parameter
$K$ and acquires different forms according to the repulsive or attractive nature of the interaction).


In this paper we focus again on the massless hypermultiplet. As in \cite{Karch:2009eb}, we notice that
at low temperature both the entropy density and the specific heat scale with the temperature as
$T^{\frac{p-3}{5-p}}$, which implies that the systems with $p\,<\,3$ have divergent entropy density and
divergent specific heat as the temperature approaches zero. Furthermore, the specific heat appears 
to be negative. For $p\,=\,3$ the system is characterised by a non-zero entropy density at zero temperature
and the specific heat scaling with the temperature depends on the codimensionality of the defect:
$c_{\mbox{\tiny $v$}}\,\sim\,T^{2(3-k)}$. Finally, for $p\,=\,4$, the entropy density and the specific
heat scales linearly in the temperature and, therefore, such a system is characterised by a zero entropy
density and a zero specific heat at zero temperature. While the thermodynamical behaviour for $p\,\ge\,3$
is physically intuitive, it is not the same for $p\,<\,3$. First, a scaling $T^{-\gamma}$ ($\gamma\,>\,0$) 
for the entropy density and specific heat seems to violate the third law of thermodynamics, which requires the
entropy density of a system to reach a minimum\footnote{Strictly speaking the third law of thermodynamics
(Nernst theorem) states that at zero temperature the entropy vanishes - provided that this limit is taken by
keeping the other thermodynamical quantities fixed. However, there exist systems which Nernst theorem
does not apply to or, more precisely, where its statement needs to be modified. For example, the previous
formulation does not apply to amorphous solids, which are not in equilibrium. They have a finite non-zero
entropy at zero temperature (holographic quantum liquids constructed with background D$3$-branes have such
a characteristic \cite{Karch:2008fa}). 
Therefore, a more general formulation of the third law of thermodynamics is that the 
entropy reach a minimum in the limit of zero temperature, which is the one we stated in the text.} 
at zero temperature (and, consequently, to have zero specific heat). 
We study the thermodynamical stability of these systems. Since we work in the 
grand-canonical ensemble, we need to check the positive-definiteness of the Hessian matrix of 
(minus) the grand-potential. The Hessian matrix turns out to be positive definite for $p\,\ge\,3$ and
negative definite for $p\,<\,3$. This is a signature of a thermodynamical instability at low temperature
for $p\,<\,3$. The natural question to ask is what is the interpretation of such an instability and how
it matches with the existence of perfectly well behaved backreacted solutions at zero temperature involving
D-branes with $p\,<\,3$, such as the Cherkis-Hashimoto solution for D$2$/D$6$ systems \cite{Cherkis:2002ir}.
We will argue that most likely this is a signature of the breaking down of the probe approximation. As 
we just mentioned, the Hessian matrix of the grand-potential turns out to be negative definite with 
just one of the eigenvalues being negative. One might think to try to stabilise
the system by turning on also a magnetic component for the world-volume gauge field strength. This however
will just produce an overall factor dependent on the magnetic charge and, as a consequence, it does not 
stabilise the system. The only tunable parameter is the number $M$ of probe branes. Increasing it, one
obtains a family of potential curves and it is possible to keep increasing the number of branes until
the Hessian changes sign. We will elaborate more extensively on this point in the Conclusion section.

The paper is organized as follows. In Section \ref{DefTh} we introduce the codimension-$k$ defect theories
by discussing their brane realisation. The codimension-$k$ defect is created by introducing probe 
D$(p+4-2k)$-brane in the background generated by the D$p$-branes in a suitable way. In Section \ref{MassHM}
we consider the probe brane embedding configuration which introduces massive fundamental degrees of
freedom in the $(p+1-k)$-dimensional defect. We explore the $\left(\mu,\,T\right)$-plane for such systems.
We show that along the chemical potential axis ($T=0$) there is a second order phase transition for any
sensible system of codimension $k\,=\,0,\,1$. We briefly discuss the case $k\,=\,2$. We also investigate
the existence of a phase transition in the interior of the $\left(\mu,\,T\right)$-plane, showing that
the order of the phase transition crucially depends on the codimensionality of the system. In Section 
\ref{hql} we consider the probe brane embedding configuration which introduces massless degrees of
freedom. We discuss the behaviour of the low temperature density entropy and specific heat. We analyse
the stability of the system by studying the positive-definiteness properties of the Hessian matrix and
we show that the systems with $p\,<\,3$ are thermodynamically unstable at low temperature.
Finally, Section \ref{Concl} contains conclusion and a summary of the results.

\section{Codimension-$k$ Defect Theories}\label{DefTh}

Consider the background generated by a stack of $N$ black D$p$-branes in the near-horizon limit
\begin{equation}\eqlabel{blackDp}
 \begin{split}
  ds_{\mbox{\tiny $10$}}^2\:&=\:
   g_{\mbox{\tiny MN}}dx^{\mbox{\tiny M}}dx^{\mbox{\tiny N}}\:=\\
  &=\:\left(\frac{r}{r_{\mbox{\tiny $p$}}}\right)^{\frac{7-p}{2}}
     \left[-h_{\mbox{\tiny $p$}}\left(r\right)dt^{2}+d\overrightarrow{x}^2\right]+
    \left(\frac{r_{\mbox{\tiny $p$}}}{r}\right)^{\frac{7-p}{2}}
     \left[\frac{dr^2}{h_{\mbox{\tiny $p$}}\left(r\right)}+r^2 \,d\Omega_{\mbox{\tiny $8-p$}}^2\right],
 \end{split}
\end{equation}
where the constant $r_{\mbox{\tiny $p$}}$ and the function  $h_{\mbox{\tiny $p$}}\left(r\right)$ are 
respectively given by
\begin{equation}\eqlabel{rp-hp}
 \begin{split}
  &r_{\mbox{\tiny $p$}}^{7-p}\:\overset{\mbox{\tiny def}}{=}\:
   \left(2\sqrt{\pi}\right)^{5-p}\Gamma\left(\frac{7-p}{2}\right) g_{\mbox{\tiny s}}N
    \left(\alpha'\right)^{\frac{7-p}{2}}\:\equiv\:
    d_{\mbox{\tiny $p$}} g_{\mbox{\tiny s}}N \left(\alpha'\right)^{\frac{7-p}{2}},\\
  &h_{\mbox{\tiny $p$}}\left(r\right)\:=\:1-\frac{r_{\mbox{\tiny $h$}}^{7-p}}{r^{7-p}},
 \end{split}
\end{equation}
with $r_{\mbox{\tiny $h$}}$ parametrising the position of the black brane horizon 
The dilaton and the background $(p+1)$-form are respectively
\begin{equation}\eqlabel{DilForm}
 e^{\phi}\:=\:
  g_{\mbox{\tiny s}}\left(\frac{r}{r_{\mbox{\tiny $p$}}}\right)^{\frac{(7-p)(p-3)}{4}},
  \quad
  C_{\mbox{\tiny $0\ldots p$}}\:=\:g_{\mbox{\tiny s}}^{-1}\left(\frac{r}{r_{\mbox{\tiny $p$}}}\right)^{7-p}.
\end{equation}
The coupling constant $g_{\mbox{\tiny YM}}$ of the dual gauge theory is dimensionful (for $p\neq3$) and 
defined by
\begin{equation}\eqlabel{gYM}
 g_{\mbox{\tiny YM}}^2\:\overset{\mbox{\tiny def}}{=}\:
  g_{\mbox{\tiny s}}\left(2\pi\right)^{p-2}\left(\alpha'\right)^{\frac{p-3}{2}}.
\end{equation}
The temperature of the background is given by the Hawking temperature
\begin{equation}\eqlabel{T}
 T\:=\:\frac{\kappa}{2\pi}\:=\:\frac{7-p}{4\pi r_{\mbox{\tiny $p$}}}
  \left(\frac{r_{\mbox{\tiny $h$}}}{r_{\mbox{\tiny $p$}}}\right)^{\frac{5-p}{2}}.
\end{equation}
Let us redefine the radial coordinate according to the following differential relation
\begin{equation}\eqlabel{rad1}
 \frac{d\sigma}{\sigma}\:=\:\frac{dr}{r\sqrt{h_{\mbox{\tiny $p$}}(r)}},
\end{equation}
so that the background metric \eqref{blackDp} takes the form
\begin{equation}\eqlabel{blackDp2}
 ds_{\mbox{\tiny $10$}}\:=\:
  \left(\frac{\sigma}{r_{\mbox{\tiny $p$}}}\right)^{\frac{7-p}{2}}\mathtt{h}_{\mbox{\tiny $+$}}
  \left[
   -\frac{\mathtt{h}_{\mbox{\tiny $-$}}^2}{\mathtt{h}_{\mbox{\tiny $+$}}^2}dt^2+
   d\overrightarrow{x}^2
  \right]+
  \left(\frac{r_{\mbox{\tiny $p$}}}{\sigma}\right)^{\frac{7-p}{2}}
   \mathtt{h}_{\mbox{\tiny $+$}}^{\frac{p-3}{7-p}}
   \left[
    d\sigma^2+\sigma^2 d\Omega_{\mbox{\tiny $8-p$}}^2
   \right],
\end{equation}
where the new radial coordinate $\sigma$ has been rescaled
\begin{equation}\eqlabel{sh}
 \sigma\:\longrightarrow\:\frac{\sigma}{\sigma_{\mbox{\tiny $h$}}},
 \qquad \sigma_{\mbox{\tiny $h$}}\:\overset{\mbox{\tiny def}}{=}\:
  \frac{r_{\mbox{\tiny $h$}}}{2^{\frac{2}{7-p}}},
 \qquad \sigma\:\in\:[\sigma_{\mbox{\tiny $h$}},\,+\infty[
\end{equation}
and the functions $\mathtt{h}_{\mbox{\tiny $\mp$}}(\sigma)$ is defined as
\begin{equation}\eqlabel{hmp}
 \mathtt{h}_{\mbox{\tiny $\mp$}}(\sigma)\:\overset{\mbox{\tiny def}}{=}\:
  1\mp\frac{\sigma_{\mbox{\tiny $h$}}^{7-p}}{\sigma^{7-p}}.
\end{equation}
The position of the black-brane horizon is now parametrized by $\sigma_{\mbox{\tiny $h$}}$ and the
background temperature can be rewritten as
\begin{equation}\eqlabel{T2}
 T\:=\:\frac{7-p}{2^{\frac{9-p}{7-p}}\pi r_{\mbox{\tiny $p$}}}
  \left(\frac{\sigma_{\mbox{\tiny $h$}}}{r_{\mbox{\tiny $p$}}}\right)^{\frac{5-p}{2}}.
\end{equation}
The dilaton and the background $(p+1)$-form respectively become
\begin{equation}\eqlabel{DilForm2}
 e^{\phi}\:=\:g_{\mbox{\tiny s}}\,\mathtt{h}_{\mbox{\tiny $+$}}^{\frac{p-3}{2}}
  \left(\frac{\sigma}{r_{\mbox{\tiny $p$}}}\right)^{\frac{(7-p)(p-3)}{4}},
 \qquad
 C_{\mbox{\tiny $0\ldots p$}}\:=\:g_{\mbox{\tiny s}}^{-1}\mathtt{h}_{\mbox{\tiny $+$}}^{2}
   \left(\frac{\sigma}{r_{\mbox{\tiny $p$}}}\right)^{7-p}.
\end{equation}

For later convenience it is also useful to express the background metric \eqref{blackDp} in a frame in which 
it is manifestly conformal to an $AdS_{\mbox{\tiny $p+2$}}\times S^{\mbox{\tiny $8-p$}}$ black-hole space 
for $p\,\neq\,5$ (the so-called ``dual'' frame)\cite{Boonstra:1998mp, Skenderis:1998dq}. 
This can be easily seen by redefining the radial coordinate according to
\begin{equation}\eqlabel{radu}
 \frac{u^2}{u_{\mbox{\tiny $p$}}^2}\:\overset{\mbox{\tiny def}}{=}\:
  \frac{r^{5-p}}{r_{\mbox{\tiny $p$}}^{7-p}},
  \qquad
  u_{\mbox{\tiny p}}\:=\:\frac{5-p}{2}
\end{equation}
and rewriting the line element \eqref{blackDp} as
\begin{equation}\eqlabel{DpConf}
  ds_{\mbox{\tiny $10$}}^{2}\:=\:\left(N\,e^{\phi}\right)^{2/(7-p)}d\hat{s}_{\mbox{\tiny $10$}}^2,
\end{equation}
so that the line element $d\hat{s}_{\mbox{\tiny $10$}}^2$ describes an $AdS_{p+2}\times S^{8-p}$ 
black-hole geometry 
\begin{equation}\eqlabel{hDp}
 \begin{split}
  ds_{\mbox{\tiny $10$}}^2\:=\:
   g_{\mbox{\tiny s}}^{-\frac{2}{7-p}}\left(\frac{r_{\mbox{\tiny $p$}}}{u_{\mbox{\tiny $p$}}}\right)^{2}
   e^{\frac{2}{7-p}\phi}
   \left\{
    u^2\left[-\mathtt{h}_{\mbox{\tiny $p$}}\left(u\right)dt^{2}+d\overrightarrow{x}^2\right]+
    \left[\mathtt{h}_{\mbox{\tiny $p$}}\left(u\right)\right]^{-1}\frac{du^2}{u^2}+
    u_{p}^2\,d\Omega_{\mbox{\tiny $8-p$}}^2
   \right\}
 \end{split}
\end{equation}
with
\begin{equation}\eqlabel{hp2}
 \mathtt{h}_{\mbox{\tiny $p$}}\left(u\right)\:=\:
  1-\left(\frac{u_{\mbox{\tiny $h$}}}{u}\right)^{2\frac{7-p}{5-p}},\qquad
  u_{\mbox{\tiny $h$}}\:=\:
  \frac{u_{\mbox{\tiny $p$}}}{r_{\mbox{\tiny $p$}}^{\frac{7-p}{2}}}\,r_{\mbox{\tiny h}}^{\frac{5-p}{2}}.
\end{equation}
Parametrising the position of the event horizon by $u_{\mbox{\tiny h}}$, the background temperature
can be conveniently written as
\begin{equation}\eqlabel{T3}
 T\:=\:\frac{7-p}{4\pi}\frac{u_{\mbox{\tiny h}}}{u_{\mbox{\tiny $p$}}}\:=\:
       \frac{7-p}{2\pi(5-p)}u_{\mbox{\tiny h}}.
\end{equation}
It is useful to redefine the radial coordinate through the following differential relation 
\begin{equation}\eqlabel{rcoord}
 \left[\mathtt{h}_{\mbox{\tiny $p$}}(u)\right]^{-\frac{1}{2}}\frac{du}{u}\:=\:
 \frac{d\rho}{\rho},
\end{equation}
so that the boundary and the black-hole horizon are now located at $\rho\,=\,0$ and $\rho\,=\,1$ respectively
and the function $\mathtt{h}_{\mbox{\tiny $p$}}(u)$ can be conveniently written as
\begin{equation}\eqlabel{hpm}
 \mathtt{h}_{\mbox{\tiny $p$}}(u)\:=\:
  \frac{\mathtt{h}_{\mbox{\tiny $-$}}^2(\rho)}{\mathtt{h}_{\mbox{\tiny $+$}}^2(\rho)},\hspace{2cm}
  \mathtt{h}_{\mbox{\tiny $\mp$}}(\rho)\:=\:1\mp\rho^{-2\frac{7-p}{5-p}}.
\end{equation}
Let us rescale the radial coordinate $\rho$ as follows
\begin{equation}\eqlabel{rhores}
 \rho\:\rightarrow\:\frac{\rho}{\rho_{\mbox{\tiny h}}}, \qquad
 \rho_{\mbox{\tiny h}}\:=\:\frac{u_{\mbox{\tiny h}}}{2^{\frac{5-p}{7-p}}},
\end{equation}
where $\rho_{\mbox{\tiny h}}$ parametrises the position of the black hole horizon. In these coordinates, 
the background metric, the dilaton and the background $(p+1)$-form becomes
\begin{equation}\eqlabel{blackDp3}
 \begin{split}
  &ds_{\mbox{\tiny $10$}}^{2}\:=\:
    \left(\frac{r_{\mbox{\tiny $p$}}}{u_{\mbox{\tiny $p$}}}\right)^{\frac{7-p}{5-p}}
    \mathtt{h}_{\mbox{\tiny $+$}}^{\frac{p-3}{7-p}}\rho^{\frac{p-3}{5-p}}
    \left\{
     \mathtt{h}_{\mbox{\tiny $+$}}^{2\frac{5-p}{7-p}}\left(\rho\right)
      \rho^2\left[-\frac{\mathtt{h}_{-}^2\left(\rho\right)}{\mathtt{h}_{+}^2\left(\rho\right)}dt^{2}+
     d\overrightarrow{x}^2\right]+
     \frac{d\rho^2}{\rho^2}+u_{\mbox{\tiny $p$}}^2 d\Omega_{\mbox{\tiny $8-p$}}^2
    \right\}\\
  &e^{\phi}\:=\:g_{\mbox{\tiny s}}
     \left(\frac{r_{\mbox{\tiny $p$}}}{u_{\mbox{\tiny $p$}}}\right)^{\frac{(p-3)(7-p)}{2(5-p)}}
     \mathtt{h}_{\mbox{\tiny $+$}}^{\frac{p-3}{2}}\rho^{\frac{(p-3)(7-p)}{2(5-p)}}
    \qquad
   C_{0\ldots p}\:=\:g_{\mbox{\tiny s}}^{-1}
    \left(\frac{r_{\mbox{\tiny $p$}}}{u_{\mbox{\tiny $p$}}}\right)^{2\frac{7-p}{5-p}}
    \mathtt{h}_{+}^2\left(\rho\right)\rho^{2\frac{7-p}{5-p}}.
 \end{split}
\end{equation}
From now on we will work with the coordinates \eqref{blackDp2}, unless otherwise specified.
\newline

In the geometry \eqref{blackDp2} let us introduce $M$ parallel probe D$(p+4-2k)$-branes $(M\ll N)$
according to the following intersection configuration
{\footnotesize
     \begin{center}
      \begin{tabular}{||p{.8cm}||*{15}{c|}|}
       \hline
             &  0  &  1  & \ldots  &  p-3  &  p-2  & p-1 &  p  & p+1 & p+2 & p+3 & p+4 & \ldots &  8  &  9  \\
       \hline
       \hline
        D$p$   &  X  &  X  & \ldots  &  X  &   X   &  X  &  X  & { } & { } & { } & { } & \ldots & { } & { } \\
       \hline
        $k=0$  &  X  &  X  & \ldots  &  X  &   X   &  X  &  X  &  X  &  X  &  X  &  X  & \ldots & {}  & {}  \\
       \hline
        $k=1$  &  X  &  X  & \ldots  &  X  &   X   &  X  & { } &  X  &  X  &  X  & { } & \ldots & { } & { } \\
       \hline
        $k=2$  &  X  &  X  & \ldots  &  X  &   X   & { } & { } &  X  &  X  & { } & { } & \ldots & { } & { } \\
       \hline
      \end{tabular}
     \end{center}
}
\noindent
intersecting the background D$p$-branes along $p-k$ directions $\left\{x^i\right\}_{i=1}^{p-k}$
and thus forming a defect of codimension $k$ ($k\,=\,0,\,1,\,2$).The probe branes wrap an internal 
$(3-k)$-sphere $S^{3-k}\subset S^{8-p}$. The transverse space can be 
parametrized as
\begin{equation}\eqlabel{dsT}
 \begin{split}
   &ds_{\mbox{\tiny T}}^2\:=\:
     d\rho^{2}+\rho^2\left(d\theta^2+\sin^2{\theta}d\Omega_{\mbox{\tiny $3-k$}}^2+
       \cos^{2}{\theta}d\Omega_{\mbox{\tiny $4-p+k$}}^2\right)\\
   &ds_{\mbox{\tiny T}}^2\:=\:d\varrho^2 + dy^2 + \varrho^2 d\Omega_{\mbox{\tiny $3-k$}}^2 + 
    y^2 d\Omega_{\mbox{\tiny $4-p+k$}}^2,
 \end{split}
\end{equation}
where the two parametrisation in \eqref{dsT} are related by 
$y=\sigma\cos{\theta},\;\varrho=\sigma\sin{\theta}$. The above configuration for the brane intersections
ensures that the systems are BPS at zero temperature and therefore no stability issues arise for the
ground state.

The presence of probe branes introduces hypermultiplets in the fundamental representation propagating on a
$\left(1+(p-k)\right)$-dimensional defect. Given the presence of a codimension-$k$ defect, the embedding of 
the D$(p+4-2k)$ branes in the D$p$-brane background can in principle be described through two functions 
$x^{p}\,\equiv\,z(\rho)$ (for $k\,\neq\,0$) and either 
$\theta\,\equiv\,\theta(\rho)$ or $y\,\equiv\,y\left(\rho\right)$ dependently on the parametrisation 
\eqref{dsT} chosen. Let us comment on these two classes of embeddings.

Fixing the position of the probe branes in the $\left((1+(p-k)\right)$-dimensional non-compact submanifold
at $z\,=\,0$, the probe branes embedding can be controlled by a scalar mode which can be turned on
by requiring that a coordinate in the transverse space has a non trivial profile. As mentioned earlier, 
the embeddings can be  parametrized either through $\theta(\rho)$ or $y(\varrho)$, dependently on the 
parametrisation of the transverse space \eqref{dsT}. 
In this class of embeddings, the space-time distance between the background
branes and the probe ones in the transverse space appears as a parameter which is related to the mass of
the fundamental hypermultiplet. Thus, analysing this class of embeddings is equivalent to consider 
massive hypermultiplets propagating in a $\left(1+(p-k)\right)$-dimensional defect.

This physical interpretation holds for $k\,=\,0,\,1$. In the case of codimension-$2$ intersections, the
gauge theory on both branes stays dynamical and can be viewed as two gauge-theories coupled through
bifundamental hypermultiplets living on a $\left(1+(p-2)\right)$-dimensional defect (the D$3$/D$3$ system
was discussed in \cite{Constable:2002xt}, while the Higgs branch for D$p$/D$p$ systems was analysed in 
\cite{Arean:2007nh}). Furthermore, the coefficient of the normalizable mode determines 
the vev of the operator dual to the embedding function, implying that the space-time separation between the 
background D$p$-branes and the probe D$p$-branes is no longer a parameter but rather provides a vev for a 
dynamical field.

The second class of embeddings describes the embedding of the $(p+2-k)$-dimensional submanifold 
$\mathcal{M}_{p+2-k}$ in the non-compact $(p+2)$-dimensional manifold $\mathcal{M}_{\mbox{\tiny $p+2$}}$
keeping the position of probe branes in the transverse space fixed to wrap the maximal sphere
$S^{\mbox{\tiny $3-k$}}$. This corresponds to set the mass of the fundamental hypermultiplet to zero.
It can be parametrized through the coordinate 
$x^{\mbox{\tiny $p$}}\,\equiv\,z\left(\rho\right)$, with $\theta\,=\,0$. The embedding mode $z$ is related
to the vev of its dual operator $\mathcal{O}_{\mbox{\tiny $z$}}$. If it is has a non-trivial profile, the
operator $\mathcal{O}_{\mbox{\tiny $z$}}$ acquires a non-zero vev breaking the supersymmetries. The 
supersymmetries are instead restored if the embedding mode is constant ($z=0$). In the case the probe
branes fill the whole non-compact manifold $\mathcal{M}_{\mbox{\tiny $p+2$}}$, obviously it is not possible
to turn on such a mode and the description of the massless excitations is necessarily supersymmetric.

For the time being, let us keep both of the two embedding functions and consider an electric component for the
world-volume gauge field strength $F_{\mbox{\tiny $AB$}}$
\begin{equation}\eqlabel{FmnAns}
 F_{2}\:=\:-f'\left(\rho\right)dt\wedge d\rho. 
\end{equation}
This means that the gauge field on the boundary field theory couples to a $U(1)\,\subset\,U(M)$ current.
With such an ansatz, the probe branes are described through the DBI-action which acquires the form
\begin{equation}\eqlabel{DprobeAct}
 \begin{split}
  S_{\mbox{\tiny $D(p+4-2k)$}}\:&=\:-M\,T_{\mbox{\tiny $D(p+4-2k)$}}\,\int d^{\mbox{\tiny p+5-2k}}\xi\:
   e^{-\phi}\sqrt{-\mbox{det}\left\{g_{\mbox{\tiny $AB$}}+F_{\mbox{\tiny $AB$}}\right\}}
  \:=\\
  &=\:-M\,T_{\mbox{\tiny $D(p+2)$}}\mathcal{N}_{\mbox{\tiny $k$}}
    \int d^{\mbox{\tiny p+2-k}}\xi\:
    \mathtt{h}_{\mbox{\tiny $-$}}\mathtt{h}_{\mbox{\tiny $+$}}^{\frac{p+1-2k}{7-p}}\varrho^{3-k}\times\\
  &\phantom{-MT}\times
    \left[
     1+\left(y'\right)^2+\left(\frac{\sigma}{r_{\mbox{\tiny $p$}}}\right)^{7-p}
      \mathtt{h}_{\mbox{\tiny $+$}}^{2\frac{5-p}{7-p}}\left(z'\right)^2-
      \frac{\mathtt{h}_{\mbox{\tiny $+$}}^{2\frac{5-p}{7-p}}}{\mathtt{h}_{\mbox{\tiny $-$}}^2}
      \left(f'\right)^2
    \right]^{\frac{1}{2}}.
   \end{split}
\end{equation}
The second line of \eqref{DprobeAct} has been written by using the solution for the dilaton \eqref{DilForm2} 
and the usual factor $\left(2\pi\alpha'\right)$ in front of the gauge field strength
has been absorbed in $F_{2}$. The induced metric on the D$(p+4-2k)$-brane world-volume metric is
\begin{equation}\eqlabel{IndMet}
 \begin{split}
  ds^2_{\mbox{\tiny $p+5-2k$}}\:=\:
   &\left(\frac{\sigma}{r_{\mbox{\tiny $p$}}}\right)^{\frac{7-p}{2}}\mathtt{h}_{\mbox{\tiny $+$}}
    \left[
     -\frac{\mathtt{h}_{\mbox{\tiny $-$}}^2}{\mathtt{h}_{\mbox{\tiny $+$}}^2}dt^2+
     d\hat{x}^2
    \right]+
    \left(\frac{r_{\mbox{\tiny $p$}}}{\sigma}\right)^{\frac{7-p}{2}}
    \mathtt{h}_{\mbox{\tiny $+$}}^{\frac{p-3}{7-p}}\times\\
   &\times
    \left[
     1+\left(y'\right)^2+\left(\frac{\sigma}{r_{\mbox{\tiny $p$}}}\right)^{7-p}
      \mathtt{h}_{\mbox{\tiny $+$}}^{2\frac{5-p}{7-p}}\left(z'\right)^2
    \right]d\varrho^2+
    \left(\frac{r_{\mbox{\tiny $p$}}}{\sigma}\right)^{\frac{7-p}{2}}
    \mathtt{h}_{\mbox{\tiny $+$}}^{\frac{p-3}{7-p}}\varrho^2 d\Omega_{\mbox{\tiny $3-k$}}^2,
 \end{split}
\end{equation}
where $\hat{x}$ indicates the (spatial) coordinates $\left\{x^i\right\}_{i=1}^{p-1}$ on the defect, the 
constant $\mathcal{N}_{\mbox{\tiny $k$}}$ is 
$\mathcal{N}_{\mbox{\tiny $k$}}\,=\,g_{\mbox{\tiny s}}^{-1}\mbox{Vol}\left\{S^{3-k}\right\}$, the 
prime $'$ indicates the first derivative with respect to the radial coordinate $\varrho$ and 
$\sigma^2\:=\:\varrho^2+y^2$. 
The action \eqref{DprobeAct} depends on the embedding function $z\left(\rho\right)$ and the gauge field 
$f\left(\rho\right)$ through their first derivatives only. There is therefore one first integral of motion 
related to each of them
\begin{equation}\eqlabel{IntMot}
 \begin{split}
  &c_{\mbox{\tiny $f$}}\:=\:\varrho^{3-k}
    \frac{\mathtt{h}_{\mbox{\tiny $+$}}^{\frac{11-p-2k}{7-p}}}{\mathtt{h}_{\mbox{\tiny $-$}}}
     \frac{-f'}{
     \sqrt{1+\left(y'\right)^2+\left(\frac{\sigma}{r_{\mbox{\tiny $p$}}}\right)^{7-p}
      \mathtt{h}_{\mbox{\tiny $+$}}^{2\frac{5-p}{7-p}}\left(z'\right)^2-
      \frac{\mathtt{h}_{\mbox{\tiny $+$}}^{2\frac{5-p}{7-p}}}{\mathtt{h}_{\mbox{\tiny $-$}}^2}
      \left(f'\right)^2}}\\
  &c_{\mbox{\tiny $z$}}\:=\:\varrho^{3-k}
    \mathtt{h}_{\mbox{\tiny $-$}}\mathtt{h}_{\mbox{\tiny $+$}}^{\frac{11-p-2k}{7-p}}
    \left(\frac{\sigma}{r_{\mbox{\tiny $p$}}}\right)^{7-p}
     \frac{z'}{
     \sqrt{1+\left(y'\right)^2+\left(\frac{\sigma}{r_{\mbox{\tiny $p$}}}\right)^{7-p}
      \mathtt{h}_{\mbox{\tiny $+$}}^{2\frac{5-p}{7-p}}\left(z'\right)^2-
      \frac{\mathtt{h}_{\mbox{\tiny $+$}}^{2\frac{5-p}{7-p}}}{\mathtt{h}_{\mbox{\tiny $-$}}^2}
      \left(f'\right)^2}}
 \end{split}
\end{equation}
Notice that, in the case of a black hole embedding phase for the class of embeddings $z$, the regularity 
condition at the horizon, fixes the first integral of motion $c_{\mbox{\tiny $z$}}$ to be zero
and therefore the embedding function $z$ must have a trivial profile. In order to have a non-trivial profile
for the embedding mode, one would need to introduce a magnetic component for the world-volume 2-form so that
the action for the probe branes has also a Wess-Zumino term. The Wess-Zumino action would induce an extra term in the 
equation of motion \eqref{IntMot}, whose value at the horizon fixes $c_{\mbox{\tiny $z$}}$.
We will not discuss the case of the presence of a magnetic component for the world-volume gauge field.
This means that the regularity condition at the horizon forces the system to be supersymmetric.

The first integral of motion $c_{\mbox{\tiny $f$}}$ is related to the charge density 
$n$, which is defined as the canonical momentum conjugate to $f\left(\rho\right)$ evaluated at the 
boundary:
\begin{equation}\eqlabel{n}
 n\:=\:\lim_{\mbox{\tiny $\varrho\rightarrow\infty$}}\frac{\partial\mathcal{L}}{\partial f'}\:=\:
  M T_{\mbox{\tiny D$(p+2)$}}\mathcal{N}_{\mbox{\tiny $k$}}\left(2\pi\alpha'\right)
  c_{\mbox{\tiny $f$}},
\end{equation}
where the factor $2\pi\alpha'$ has been restored. For later convenience, let us rescale the coordinates, the
gauge field $f$ and $c_{\mbox{\tiny $f$}}$ so that they are dimensionless:
\begin{equation}\eqlabel{scale}
 \left\{t,\,\overrightarrow{x},\,\varrho,\,y;\,f\right\}\:\rightarrow\:
 L_{\mbox{\tiny $\star$}}\left\{t,\,\overrightarrow{x},\,\varrho,\,y;\,f\right\},\qquad
 c_{\mbox{\tiny $f$}}\:\rightarrow\:L_{\mbox{\tiny $\star$}}^{3-k}c_{\mbox{\tiny $f$}}.
\end{equation}
With such a rescaling, the position of the black brane horizon gets parametrized by the dimensionless
quantity $\hat{\sigma}_{\mbox{\tiny $h$}}\,=\,\sigma_{\mbox{\tiny $h$}}/L_{\mbox{\tiny $\star$}}$.

We will discuss the two classes of embeddings separately, 
{\it i.e.} we will discuss both the massive and massless degrees of freedom.

\section{Massive hypermultiplet and $(\mu,\,T)$ phase diagram}\label{MassHM}

Let us now fix the position of the probe D$(p+4-2k)$-branes in the non-compact 
$\left(1+(p-k)\right)$-dimensional submanifold ($z=0$) and let us consider their embedding in the transverse 
space, which has been parametrized 
through the angular coordinate $\theta\left(\rho\right)$. It is actually more convenient to parametrise 
differently the embedding, using a function $y\left(\varrho\right)$ of a redefined radial coordinate $\varrho$
according to \eqref{blackDp2} and \eqref{dsT}. Let us write explicitly the induced metric on the 
world-volume of the probe branes
\begin{equation}\eqlabel{IndMet2}
 ds^{2}_{\mbox{\tiny $p+5-2k$}}\:=\:
  \left(\frac{\sigma}{r_{\mbox{\tiny $p$}}}\right)^{\frac{7-p}{2}}
  \mathtt{h}_{\mbox{\tiny $+$}}
  \left[
   -\frac{\mathtt{h}_{\mbox{\tiny $-$}}^2}{\mathtt{h}_{\mbox{\tiny $+$}}^2}
   +d\hat{x}^2
  \right]+
  \left(\frac{r_{\mbox{\tiny $p$}}}{\sigma}\right)^{\frac{7-p}{2}}
  \mathtt{h}_{\mbox{\tiny $+$}}^{\frac{p-3}{7-p}}
  \left[
   \left(1+\left(y'\right)^2\right)d\varrho^2 + \varrho^2 d\Omega_{\mbox{\tiny $3-k$}}^2
  \right],
\end{equation}
with $\left\{x^i\right\}_{\mbox{\tiny $i=1$}}^{\mbox{\tiny $p-k$}}$, 
$\:\sigma^2\:=\:\varrho^2+y^2$ and the dilaton given by
\begin{equation}\eqlabel{Dil3}
 e^{\phi}\:=\:g_{\mbox{\tiny s}}\mathtt{h}_{\mbox{\tiny $+$}}^{\frac{p-3}{2}}
   \left(\frac{\sigma}{r_{\mbox{\tiny $p$}}}\right)^{\frac{(7-p)(p-3)}{4}},
 \qquad
 \mathtt{h}_{\mbox{\tiny $\mp$}}\:=\:1\mp\left(\frac{\hat{\sigma}_{\mbox{\tiny $h$}}}{\sigma}\right)^{7-p},
\end{equation}
where the coordinates are dimensionless. For this class of embeddings, the action of the probe branes is given
just by the DBI-action
\begin{equation}\eqlabel{DBIactionY}
 \begin{split}
 S_{\mbox{\tiny D$(p+4-2k)$}}\:=\:-MT_{\mbox{\tiny D$(p+4-2k)$}}\mathcal{N}_{\mbox{\tiny $k$}}
  \int d^{p-k+2}\xi\:
   \mathtt{h}_{\mbox{\tiny $-$}}\mathtt{h}_{\mbox{\tiny $+$}}^{\frac{p+1-2k}{7-p}}\varrho^{3-k}
   \left[1+\left(y'\right)^2-
    \frac{\mathtt{h}_{\mbox{\tiny $+$}}^{2\frac{5-p}{7-p}}}{\mathtt{h}_{\mbox{\tiny $-$}}^2}
    \left(f'\left(\varrho\right)\right)^2\right]^{\frac{1}{2}}.
 \end{split}
\end{equation}
As noticed earlier, in the most general case, the action \eqref{DBIactionY} depends on the electric gauge 
potential $f\left(\varrho\right)$ just through its first derivative, so that there exists a first integral of 
motion $c_{\mbox{\tiny $f$}}$ related to it. In the limit of zero temperature (i.e. in any point of the 
chemical potential axis in the phase diagram $\left(\mu,\,T\right)$), the action also depends on the embedding
function $y$ through its first derivative only, so that there is a further integral of motion 
$c_{\mbox{\tiny $y$}}$, beside $c_{\mbox{\tiny $f$}}$. In the most general case, the equations of motion are
\begin{equation}\eqlabel{eomYF}
 \begin{split}
  &f'\left(\varrho\right)\:=\:
    c_{\mbox{\tiny $f$}}\frac{\mathtt{h}_{\mbox{\tiny $-$}}}{\mathtt{h}_{\mbox{\tiny $+$}}^{\frac{5-p}{7-p}}}
     \frac{\sqrt{1+\left(y'\right)^2}}{\sqrt{{c_{\mbox{\tiny $f$}}^2+\varrho^{3-k}}
      \mathtt{h}_{\mbox{\tiny $+$}}^{4\frac{3-k}{7-p}}}},\\
  &0\:=\:\frac{y''}{1+\left(y'\right)^2}+\frac{3-k}{\varrho}y'+
    2\left(\frac{\hat{\sigma}_{\mbox{\tiny h}}}{\sigma}\right)^{7-p}
    \frac{\varrho y'-y}{\sigma^2 \mathtt{h}_{\mbox{\tiny $-$}}  \mathtt{h}_{\mbox{\tiny $+$}}}
    \left[3+k-p+\left(4-k\right)\left(\frac{\hat{\sigma}_{\mbox{\tiny h}}}{\sigma}\right)^{7-p}\right]-\\
  &\phantom{0\:=\:}
    -\frac{c_{\mbox{\tiny $f$}}^2}{c_{\mbox{\tiny $f$}}^2+
     \varrho^{2(3-k)}\mathtt{h}_{\mbox{\tiny $+$}}^{4\frac{3-k}{7-p}}}
    \left[
     \frac{3-k}{\varrho}y'-2\left(3-k\right)\left(\frac{\hat{\sigma}_{\mbox{\tiny h}}}{\sigma}\right)^{7-p}
       \frac{\varrho y'-y}{\sigma^2 \mathtt{h}_{\mbox{\tiny $+$}}}
    \right]
 \end{split}
\end{equation}
In the following subsections, we explore the whole phase diagram for this class of systems, including
also the simplest cases.

\subsection{The chemical potential axis}\label{cpaxis}

As mentioned earlier, in the limit of zero temperature  there is a first integral
of motion $c_{\mbox{\tiny $y$}}$ related to the embedding function $y$, beside $c_{\mbox{\tiny $f$}}$.
The equations of motion \eqref{eomYF} have the following simple form
\begin{equation}\eqlabel{eomYFa}
  f'\left(\varrho\right)\:=\:\frac{c_{\mbox{\tiny $f$}}}{
    \sqrt{\varrho^{2(3-k)}+c_{\mbox{\tiny $f$}}^2-c_{\mbox{\tiny $y$}}^2}},\qquad
  y'\left(\varrho\right)\:=\:\frac{c_{\mbox{\tiny $y$}}}{\sqrt{\varrho^{2(3-k)}+c_{\mbox{\tiny $f$}}^2-
    c_{\mbox{\tiny $y$}}^2}}.
\end{equation}
They can be easily integrated out. Notice that we need to distinguish three cases, according to the
sign of $c_{\mbox{\tiny $f$}}^2-c_{\mbox{\tiny $y$}}^2$. In this case this difference is negative, 
the probe branes can extend from the boundary to a minimum distance $\varrho_{\mbox{\tiny min}}\,=\,
\left(c_{\mbox{\tiny $y$}}^2-c_{\mbox{\tiny $f$}}^2\right)^{1/2(3-k)}$, where the branes turns back and
hit the boundary again. The system is therefore a D$(p+4-2k)$/$\bar{\mbox{D}}(p+4-2k)$ in a D$p$-brane 
background. In this phase, the chemical potential is
\begin{equation}\eqlabel{cp1}
 \mu\:=\:\int_{\varrho_{\mbox{\tiny min}}}^{\infty}d\varrho\:\frac{c_{\mbox{\tiny $f$}}}{
    \sqrt{\varrho^{2(3-k)}+c_{\mbox{\tiny $f$}}^2-c_{\mbox{\tiny $y$}}^2}}\:=\:
    \frac{1}{2(3-k)}\frac{c_{\mbox{\tiny $f$}}}{\left(c_{\mbox{\tiny $y$}}^2-c_{\mbox{\tiny $f$}}^2
    \right)^{\frac{2-k}{2(3-k)}}}B\left(\frac{2-k}{2(3-k)},\frac{1}{2}\right),
\end{equation}
while the embedding function can be explicitly written as
\begin{equation}\eqlabel{ef1}
 \begin{split}
  y\left(\varrho\right)\:&=\:\int_{\varrho_{\mbox{\tiny min}}}^{\varrho}d\varrho'\:
     \frac{c_{\mbox{\tiny $y$}}}{\sqrt{\varrho'^{2(3-k)}+c_{\mbox{\tiny $f$}}^2-c_{\mbox{\tiny $y$}}^2}}\:=\:\\
  &=\:\frac{1}{4}\frac{c_{\mbox{\tiny $y$}}}{\left(c_{\mbox{\tiny $y$}}^2-c_{\mbox{\tiny $f$}}^2
     \right)^{\frac{2-k}{2(3-k)}}}\left[B\left(\frac{2-k}{2(3-k)},\frac{1}{2}\right)-
     B\left(\left(\frac{\varrho_{\mbox{\tiny min}}}{\varrho}\right)^{\frac{2-k}{2(3-k)}};\frac{2-k}{2(3-k)},
      \frac{1}{2}\right)\right].
 \end{split}
\end{equation}
In \eqref{cp1} and \eqref{ef1}, we set 
$f\left(\varrho_{\mbox{\tiny min}}\right)\,=\,0\,=\,y\left(\varrho_{\mbox{\tiny min}}\right)$. 
Notice that the expressions \eqref{cp1} and \eqref{ef1} for the chemical potential and the embedding 
functions are valid for $k\,=\,0,\,1$. For the codimension-$2$ systems, the integrals \eqref{cp1} and 
\eqref{ef1} are divergent as $\varrho\,\rightarrow\,\infty$. We will come back to this issue later. For
the time being, we focus on the case $k\,=\,0,\,1$.

Taking the limit $\varrho\,\rightarrow\,\infty$ of \eqref{ef1} (in this limit the incomplete Beta-function
vanishes), one obtains the ``quark'' mass $m$ in terms of the first integral of motions $c_{\mbox{\tiny $f$}}$
and $c_{\mbox{\tiny $y$}}$. It is easy to invert this expression, together with the result for the
chemical potential \eqref{cp1}, to obtain the first integral of motions in terms of $m$ and $\mu$:
\begin{equation}\eqlabel{cfcy1}
 c_{\mbox{\tiny $f$}}\:=\:
  \left[\frac{2(3-k)}{B\left(\frac{2-k}{2(3-k)},\frac{1}{2}\right)}\right]^{3-k}
  \mu\left(m^2-\mu^2\right)^{\frac{2-k}{2}},\quad
 c_{\mbox{\tiny $y$}}\:=\:
  \left[\frac{2(3-k)}{B\left(\frac{2-k}{2(3-k)},\frac{1}{2}\right)}\right]^{3-k}
  m\left(m^2-\mu^2\right)^{\frac{2-k}{2}}.
\end{equation}
The condition $c_{\mbox{\tiny $f$}}^2- c_{\mbox{\tiny $y$}}^2\,<\,0$ implies necessarily that
$m\,>\,\mu$ for $k$ even. For $k\,=\,1$, it is straightforward to realize that the expressions
\eqref{cfcy1} are meaningful just for in the region $m\,>\,\mu$ (the region $m\,<\,-\mu$ is not
physical).

In the case $c_{\mbox{\tiny $f$}}^2-c_{\mbox{\tiny $y$}}^2\,>\,0$, the probe branes can extend
to $\varrho\,=\,0$. Such a configuration corresponds to a black-hole crossing phase, in which
the chemical potential and the embedding function are expressed by
\begin{equation}\eqlabel{cpef2}
 \begin{split}
  &\mu\:=\:\int_{0}^{\infty}d\varrho\:\frac{c_{\mbox{\tiny $f$}}}{\sqrt{\varrho^{2(3-k)}+
    c_{\mbox{\tiny $f$}}^2-c_{\mbox{\tiny $y$}}^2}}\:=\:
    \frac{1}{2(3-k)}\frac{c_{\mbox{\tiny $f$}}}{
     \left(c_{\mbox{\tiny $f$}}^2-c_{\mbox{\tiny $y$}}^2\right)^{\frac{2-k}{2(3-k)}}}\,
     B\left(\frac{2-k}{2(3-k)},\frac{1}{2(3-k)}\right),\\
  &y\left(\rho\right)\:=\:\frac{1}{2(3-k)}\frac{c_{\mbox{\tiny $y$}}}{
     (c_{\mbox{\tiny $f$}}^2-c_{\mbox{\tiny $y$}}^2)^{\frac{2-k}{2(3-k)}}}\,
     \left[
      B\left(\frac{2-k}{2(3-k)},\,\frac{1}{2(3-k)}\right)-
     \right.\\
    &\left.\phantom{y\left(\rho\right)\:=\:}\:-
      B\left(\frac{c_{\mbox{\tiny $f$}}^2-c_{\mbox{\tiny $y$}}^2}{\varrho^{2(3-k)}};\;
       \frac{2-k}{2(3-k)},\frac{1}{2(3-k)}\right)
     \right].
 \end{split}
\end{equation}
At the boundary, the embedding function provides the ``quark'' mass, which is provided by the 
first term in the second expression in \eqref{cpef2} (the incomplete Beta-function vanishes).
The first integrals of motion $c_{\mbox{\tiny $f$}}$ and $c_{\mbox{\tiny $y$}}$ can be easily expressed
in terms of $m$ and $\mu$:
\begin{equation}\eqlabel{cfcy2}
 c_{\mbox{\tiny $f$}}\:=\:
  \left[
   \frac{2(3-k)}{B\left(\frac{2-k}{2(3-k)},\,\frac{1}{2(3-k)}\right)}
  \right]^{3-k}
  \mu\left(\mu^2-m^2\right)^{\frac{2-k}{2}},\quad
  c_{\mbox{\tiny $y$}}\:=\:
  \left[
   \frac{2(3-k)}{B\left(\frac{2-k}{2(3-k)},\,\frac{1}{2(3-k)}\right)}
  \right]^{3-k}
  m\left(\mu^2-m^2\right)^{\frac{2-k}{2}}.
\end{equation}
Similarly to the previous case, it is straightforward to see that equations \eqref{cfcy2} hold
in the phase diagram region $\mu\,>\,m$.


Let us now turn to the thermodynamics of our systems in these two phases, by considering in particular
the grand-potential $\Omega$. It is given, up to a sign, by the renormalized on-shell 
action\footnote{The holographic renormalization of probe D-branes in non-conformal background was extensively 
discussed in \cite{Benincasa:2009ze}.}
\begin{equation}\eqlabel{GP1}
 \begin{split}
  &\Omega\:=\:-S_{\mbox{\tiny D$(p+4-2k)$}}\Big|_{\mbox{\tiny ren}}\:=\:
             -\lim_{\Lambda\rightarrow\infty}
             \left[
              S_{\mbox{\tiny D$(p+4-2k)$}}\Big|_{\mbox{\tiny on-shell}}+
              S_{\mbox{\tiny D$(p+4-2k)$}}\Big|_{\mbox{\tiny ct}}
             \right]
 \end{split}
\end{equation}
where
\begin{equation}\eqlabel{Sren}
 \begin{split}
  &S_{\mbox{\tiny D$(p+4-2k)$}}\Big|_{\mbox{\tiny on-shell}}\:=\:
   -M T_{\mbox{\tiny D$(p+4-2k)$}}\hat{\mathcal{N}}_{\mbox{\tiny $k$}}
     \int_{\tilde{\varrho}}^{\Lambda}d\varrho\:
     \frac{\varrho^{2(3-k)}}{\sqrt{\varrho^{2(3-k)}+c_{\mbox{\tiny $f$}}^2-c_{\mbox{\tiny $y$}}^2}}\\
  &S_{\mbox{\tiny D$(p+4-2k)$}}\Big|_{\mbox{\tiny ct}}\:=\:
    -M T_{\mbox{\tiny D$(p+4-2k)$}}\hat{\mathcal{N}}_{\mbox{\tiny $k$}}
    \left(-\frac{\Lambda^{4-k}}{4-k}\right),
 \end{split}
\end{equation}
(the overall factor $\hat{\mathcal{N}}_{\mbox{\tiny $k$}}$ is a redefinition of 
$\mathcal{N}_{\mbox{\tiny $k$}}$ including 
$\int d^{\mbox{\tiny $p+k-1$}}\xi$, and $\tilde{\varrho}$ is
0 for $c_{\mbox{\tiny $f$}}^2-c_{\mbox{\tiny $y$}}^2\,>\,0$ or $\varrho_{\mbox{\tiny min}}$ if
$c_{\mbox{\tiny $f$}}^2-c_{\mbox{\tiny $y$}}^2\,<\,0$).

Integrating the expressions \eqref{Sren}, the grand-potential in the two phases is given by
\begin{equation}\eqlabel{gp1}
 \Omega\:=\:
  \left\{
   \begin{array}{l}
    -\frac{\left(c_{\mbox{\tiny $f$}}^2-c_{\mbox{\tiny $y$}}^2\right)^{\frac{(4-k)}{2(3-k)}}}{2(4-k)}
     B\left(\frac{2-k}{2(3-k)},\,\frac{7-2k}{2-k}\right)\\
    \phantom{\ldots}\\
    \frac{\left(c_{\mbox{\tiny $y$}}^2-c_{\mbox{\tiny $f$}}^2\right)^{\frac{(4-k)}{2(3-k)}}}{2(3-k)(4-k)}
     B\left(\frac{2-k}{2(3-k)},\,\frac{1}{2}\right)
   \end{array}
  \right.
  \:=\:
  \left\{
   \begin{array}{l}
    \phantom{\frac{(1^{2})^{\frac{1}{2}}}{2}}\hspace{-0.8cm}
     -\mathtt{a}_{1}\left(\mu^2-m^2\right)^{\frac{4-k}{2}},\qquad \mu\,>\,m\\	
    \phantom{\ldots}\\
    \phantom{\frac{(1^{2})^{\frac{1}{2}}}{2}}\hspace{-0.8cm}
     \mathtt{a}_{2}\left(m^2-\mu^2\right)^{\frac{4-k}{2}}, \qquad \mu\,<\,m
   \end{array}
  \right.
\end{equation}
where $\mathtt{a}_{1}$ and $\mathtt{a}_{2}$ are two positive constants. 

Some comments are now in order. Notice that the phase (``black-hole'' crossing phase) for which 
$c_{\mbox{\tiny $f$}}^2-c_{\mbox{\tiny $y$}}^2\,>\,0$ is thermodynamically favoured with respect to the
phase (``brane/anti-brane''-phase) for which this difference is negative, given that in the first case the
grand-potential is negative while in the second case it is positive. Let us analyze the derivatives
of the grand-potential $\Omega$:
\begin{equation}\eqlabel{GPder1}
 \begin{split}
  &\left.\frac{\partial\Omega}{\partial\mu}\right|_{\mbox{\tiny $\mu\rightarrow m$}}\:=\:
   \left\{
    \begin{array}{l}
     \left.-\mathtt{a}_{\mbox{\tiny $1$}}(4-k)\mu\left(\mu^2-m^2\right)^{\frac{2-k}{2}}
     \right|_{\mbox{\tiny $\mu\rightarrow m$}}\\
     \phantom{\ldots}\\
     \left.-\mathtt{a}_{\mbox{\tiny $2$}}(4-k)\mu\left(m^2-\mu^2\right)^{\frac{2-k}{2}}
     \right|_{\mbox{\tiny $\mu\rightarrow m$}}
    \end{array}
   \right.
   \:\propto\:
   \left. c_{\mbox{\tiny $f$}}\right|_{\mbox{\tiny $\mu\rightarrow m$}}\:=\:0\\
  &\left.\frac{\partial^2\Omega}{\partial\mu^2}\right|_{\mbox{\tiny $\mu\rightarrow m$}}\:=\:
   \left\{
    \begin{array}{l}
     \left.-\mathtt{a}_{\mbox{\tiny $1$}}(4-k)\frac{(3-k)\mu^2-m^2}{\left(\mu^2-m^2\right)^{\frac{k}{2}}}
     \right|_{\mbox{\tiny $\mu\rightarrow m$}}\\
     \phantom{\ldots}\\
     \left.\mathtt{a}_{\mbox{\tiny $2$}}(4-k)\frac{(3-k)\mu^2-m^2}{\left(\mu^2-m^2\right)^{\frac{k}{2}}}
     \right|_{\mbox{\tiny $\mu\rightarrow m$}}
    \end{array}
   \right.
   \:\propto\:
   \left.\frac{\partial c_{\mbox{\tiny $f$}}}{\partial\mu}\right|_{\mbox{\tiny $\mu\rightarrow m$}}
   \:\rightarrow\:\infty
 \end{split}
\end{equation}
The point $\mu\:=\:m$ represents a second order phase transition. This is a universal feature for gauge 
theories whose holographic bulk dual is constructed in terms of flavoured branes. In this section we showed
how this holds for gauge theories with no defect ($k\,=\,0$) and on a defect with $(p-1)$ spatial directions 
($k=1$). The presence of this second order phase transition is however common to gauge theories on a defect 
with $(p-2)$ spatial directions as well, which is dual to the other possible BPS brane construction 
($k\,=\,2$). In this last case one needs to take into account some subtleties which we will discuss in the 
next subsection.

\subsubsection{Codimension-2 systems}

As pointed out, the analysis in the previous subsection holds for codimension-$0$ and codimension-$1$ systems,
{\it i.e. } D$p$/D$(p+4)$ and D$p$/D$(p+2)$ intersections respectively. The D$p$/D$p$ systems are quite	
different from the higher dimensional defect theories. The first difference lies in the divergence structure
of the on-shell action\footnote{For the discussion of the D$3$/D$3$ system see section 6 of 
\cite{Karch:2005ms}. The generalisation to any D$p$/D$p$ system ($p<5$) was discussed in 
\cite{Benincasa:2009ze}}. 
The case we are now analysing is the only one in which the embedding mode saturates
the Breitenlhoner-Freedman bound which implies the presence of a logarithmic divergence
\cite{Karch:2005ms, Benincasa:2009ze}. Furthermore, the role of the normalizable and non-normalizable modes
are exchanged, with the coefficient of the normalizable mode determining the vacuum-expectation-value of the
operator dual to the embedding function, so that the brane separation appears as a vev rather than as
a parameter.
A similar discussion applies for the gauge potential $f(\rho)$. For both $y(\rho)$ and $f(\rho)$,
the asymptotic expansion near the boundary $\rho\,\rightarrow\,\infty$ shows a logarithmic term
\begin{equation}\eqlabel{yfasymp}
 y(\rho)\:\overset{\mbox{\tiny $\rho\rightarrow\infty$}}{=}\:m+c_{\mbox{\tiny $y$}}\,\log{\rho},\qquad
 f(\rho)\:\overset{\mbox{\tiny $\rho\rightarrow\infty$}}{=}\:\mu+c_{\mbox{\tiny $f$}}\,\log{\rho}.
\end{equation}
The divergences of the action \eqref{Sren} are cured by the counter-terms found in \cite{Benincasa:2009ze}.
In principle, the integrals of the equations of motion \eqref{eomYFa}, which define $y(\rho)$ and $f(\rho)$, 
remain divergent as near the boundary. One can define $\mu$ and $m$ respectively as
\begin{equation}\eqlabel{mum}
 \mu\:=\:-\frac{\partial\Omega}{\partial c_{\mbox{\tiny $f$}}},\qquad
 m\:=\:\frac{\partial\Omega}{\partial c_{\mbox{\tiny $y$}}}.
\end{equation}
Through \eqref{mum}, one can rewrite the grand-potential in the form \eqref{gp1} (with $k\,=\,2$). Therefore, 
codimension-$2$ systems show a second order phase transition on the chemical potential axis at $\mu\,=\,m$
as well. 

Thus one can conclude that the existence of a second order phase transition on the chemical potential
axis is a universal feature of gauge theories with a gravitational bulk dual constructed via D$p$/D$(p+4-2k)$
systems.

Before turning on the temperature, let us discuss the behaviour of the chemical potential and of the embedding
function in the black-hole crossing phase in the limit of small density $c_{\mbox{\tiny $f$}}$. This will
turn out to be useful for the finite temperature case, where the analytic analysis is performed in such a
limit.

\subsubsection{Small density expansion}

Consider the expression for the embedding function \eqref{ef1} taking the limit 
$\varrho\,\rightarrow\,\infty$, so that it provides an expression of the ``quark''-mass in terms of the two
first integral of motion $c_{\mbox{\tiny $f$}}$ and $c_{\mbox{\tiny$y$}}$. Such an expression  
can be inverted in order to express $c_{\mbox{\tiny$y$}}$ in terms of the density $c_{\mbox{\tiny $f$}}$.
In the limit  $c_{\mbox{\tiny $f$}}\,\rightarrow\,0$, one obtains:
\begin{equation}\eqlabel{cysmallcf}
 c_{\mbox{\tiny $f$}}^2-c_{\mbox{\tiny $y$}}^2\:=\:
  \kappa^{2\frac{3-k}{2-k}}c_{\mbox{\tiny $f$}}^{2\frac{3-k}{2-k}}
  \left[
   1-\frac{3-k}{2-k}\kappa^{2\frac{3-k}{2-k}}c_{\mbox{\tiny $f$}}^{\frac{2}{2-k}}+
    \mathcal{O}\left(c_{\mbox{\tiny $f$}}^{\frac{4}{2-k}}\right)
  \right],
\end{equation}
where $\kappa$ is constant which contains the ``quark'' mass as well as other numerical coefficients. 

This suggests that a perturbative expansion around $c_{\mbox{\tiny $f$}}$ becomes subtle in the region
of the radial axis for which $\varrho^{2(3-k)}$ is of the same order of the leading term in 
\eqref{cysmallcf}, i.e.
\begin{equation}\eqlabel{rr}
 \varrho\:\sim\:c_{\mbox{\tiny $f$}}^{\frac{1}{2-k}}.
\end{equation}
Therefore, this region of the radial axis can be conveniently studied by redefining the radial coordinate
as $\tau\,=\,c_{\mbox{\tiny $f$}}^{\frac{1}{2-k}}\varrho$. Keeping this in mind, it is possible to integrate
the equation for the gauge field \eqref{eomYFa} to obtain the chemical potential in the small density limit
at zero temperature
\begin{equation}\eqlabel{mu0small}
 \mu\:=\:m_{\mbox{\tiny $0$}}\left[1+\frac{\kappa^{2\frac{3-k}{2-k}}}{2}c_{\mbox{\tiny $f$}}^{\frac{2}{2-k}}+
         \mathcal{O}\left(c_{\mbox{\tiny $f$}}^{\frac{4}{2-k}}\right)\right].
\end{equation}
Given equation \eqref{mu0small}, one can expect that the expansion parameter for the chemical potential 
(as well as the embedding function) is $c_{\mbox{\tiny $f$}}^{\frac{2}{2-k}}$ rather than other powers
of the ``quark''-density $c_{\mbox{\tiny $f$}}$.


\subsection{Chemical potential-temperature plane}\label{Cpsde}

In this section we consider the systems at finite temperature and with chemical potential. For convenience we 
rewrite the equations of motions here
\begin{equation}\eqlabel{eqomYF2}
 \begin{split}
  &f'\left(\varrho\right)\:=\:
   c_{\mbox{\tiny $f$}}\frac{\mathtt{h}_{\mbox{\tiny $-$}}}{\mathtt{h}_{\mbox{\tiny $+$}}^{\frac{5-p}{7-p}}}
   \frac{\sqrt{1+\left(y'\right)^2}}{\sqrt{c_{\mbox{\tiny $f$}}^2+
    \varrho^{2(3-k)}\mathtt{h}_{\mbox{\tiny $+$}}^{4\frac{3-k}{7-p}}}}\\
   &0\:=\:\frac{y''}{1+\left(y'\right)^2}+\frac{3-k}{\varrho}y'+
    2\left(\frac{\hat{\sigma}_{\mbox{\tiny h}}}{\sigma}\right)^{7-p}
    \frac{\varrho y'-y}{\sigma^2 \mathtt{h}_{\mbox{\tiny $-$}}  \mathtt{h}_{\mbox{\tiny $+$}}}
    \left[3+k-p+\left(4-k\right)\left(\frac{\hat{\sigma}_{\mbox{\tiny h}}}{\sigma}\right)^{7-p}\right]-\\
  &\phantom{0\:=\:}
    -\frac{c_{\mbox{\tiny $f$}}^2}{c_{\mbox{\tiny $f$}}^2+
     \varrho^{2(3-k)}\mathtt{h}_{\mbox{\tiny $+$}}^{4\frac{3-k}{7-p}}}
    \left[
     \frac{3-k}{\varrho}y'-2\left(3-k\right)\left(\frac{\hat{\sigma}_{\mbox{\tiny h}}}{\sigma}\right)^{7-p}
       \frac{\varrho y'-y}{\sigma^2 \mathtt{h}_{\mbox{\tiny $+$}}}
    \right]
  \end{split}
\end{equation}
In the black-hole crossing phase, the regularity condition at the horizon imposes the following boundary 
conditions for the embedding function
\begin{equation}\eqlabel{bhbc}
 y'\left(\varrho_{\mbox{\tiny h}}\right)\:=\:\frac{y_{\mbox{\tiny h}}}{\varrho_{\mbox{\tiny h}}}
 \qquad
 \mbox{with }
 \sigma_{\mbox{\tiny h}}^2\:=\:\varrho_{\mbox{\tiny h}}^2+y_{\mbox{\tiny h}}^2,
\end{equation}
which prevents the last two terms in \eqref{eqomYF2} to blow-up as the horizon is approached.
From the first equation in \eqref{eqomYF2}, the chemical potential can be written as
\begin{equation}\eqlabel{cp}
\mu\:=\:
  c_{\mbox{\tiny $f$}}\int_{\varrho_{\mbox{\tiny h}}}^{\infty}d\varrho\:
   \frac{\mathtt{h}_{\mbox{\tiny $-$}}}{\mathtt{h}_{\mbox{\tiny $+$}}^{\frac{5-p}{7-p}}}
   \frac{\sqrt{1+\left(y'\right)^2}}{\sqrt{c_{\mbox{\tiny $f$}}^2+\varrho^{2(3-k)}
   \mathtt{h}_{\mbox{\tiny $+$}}^{4\frac{3-k}{7-p}}}}.
\end{equation}
In order to study the transition line in the $\left(\mu,\,T\right)$-plane analytically, we closely follow
the small density approach of \cite{Faulkner:2008hm}.

\subsubsection{Small density expansion of the embedding function}

As noted in \cite{Faulkner:2008hm}, the small density expansion for the embedding function and the
chemical potential is subtle. This is due to the fact that the last term in the embedding function equation
\eqref{eqomYF2} for $c_{\mbox{\tiny $f$}}\rightarrow0$ can be considered as an actual perturbation of
the equation at zero condensate as long as $\varrho^{2(3-k)}\mathtt{h}_{\mbox{\tiny $+$}}^{4\frac{3-k}{7-p}}$ 
does not become of order $\mathcal{O}\left(c_{\mbox{\tiny $f$}}^2\right)$. For such values of the radial coordinates, 
this term is no longer small. Let us divide the radial direction in two regions:
\begin{enumerate}
 \item $\varrho\,\in\,]\varrho_{\mbox{\tiny $\Lambda$}},\,\infty[$, \\
       where $\varrho_{\mbox{\tiny $\Lambda$}}$ is a cut-off distance until which a uniform perturbative
       expansion can be considered;
 \item $\tau\,\in\,]\tau_{\mbox{\tiny h}},\,\Lambda[$\;,\\
       where $\tau$ is defined as 
       $\varrho\:=\:c_{\mbox{\tiny $f$}}^{\frac{1}{2-k}}\tau$, and $\tau_{\mbox{\tiny h}}$ and $\Lambda$ 
       respectively as 
       $\varrho_{\mbox{\tiny h}}\:=\:c_{\mbox{\tiny $f$}}^{\frac{1}{2-k}}\tau_{\mbox{\tiny h}}$ and
       $\varrho_{\mbox{\tiny $\Lambda$}}\:=\:c_{\mbox{\tiny $f$}}\Lambda$.
\end{enumerate}
The idea is to found a (perturbative) solution in both regions 1 and 2 and then matching them in the 
limit $c_{\mbox{\tiny $f$}}\rightarrow0$, $\varrho_{\mbox{\tiny $\Lambda$}}\rightarrow0$ and
$\Lambda\rightarrow\infty$, keeping $\tau$ finite. Notice that the above splitting of the radial
coordinate axis does not hold for $k\,=\,2$.
i
Let us start with finding the solution for the embedding function equation \eqref{eqomYF2} in the
region 1, where, for small $c_{\mbox{\tiny $f$}}$, the solution can be expanded as follows
\begin{equation}\eqlabel{yexp}
 y\left(\varrho\right)\:=\:\sum_{i=0}^{\infty}
  c_{\mbox{\tiny $f$}}^{i}
  y_{\mbox{\tiny $i$}}\left(\varrho\right).
\end{equation}
The zero-th order term $y_{\mbox{\tiny $0$}}\left(\varrho\right)$ satisfied the equation \eqref{eqomYF2} at
$c_{\mbox{\tiny $f$}}=0$ with boundary condition $y_{\mbox{\tiny $0$}}\left(0\right)=1$. In
a small $\varrho$ expansion, the 0-th order (in $c_{\mbox{\tiny $f$}}$) solution is given by
\begin{equation}\eqlabel{y0r}
 y_{\mbox{\tiny $0$}}\left(\varrho\right)\:=\:1+y_{\mbox{\tiny $0$}}^{\mbox{\tiny $(2)$}}\rho^2+
  y_{\mbox{\tiny $0$}}^{\mbox{\tiny $(4)$}}\rho^4+\mathcal{O}\left(\varrho^6\right),
\end{equation}
where the coefficients $y_{\mbox{\tiny $0$}}^{\mbox{\tiny $(2)$}}$, 
$y_{\mbox{\tiny $0$}}^{\mbox{\tiny $(4)$}}$ and $y_{\mbox{\tiny $0$}}^{\mbox{\tiny $(6)$}}$ are given by
\begin{equation}\eqlabel{y0rCoeffs}
 \begin{split}
   &y_{\mbox{\tiny $0$}}^{\mbox{\tiny $(2)$}}\:=\:
    \frac{(3+k-p)+(4-k)\hat{\sigma}_{\mbox{\tiny h}}^{7-p}}{(4-k)
     \left(1-\hat{\sigma}_{\mbox{\tiny h}}^{2(7-p)}\right)}\,\hat{\sigma}_{\mbox{\tiny h}}^{7-p},\\
   &y_{\mbox{\tiny $0$}}^{\mbox{\tiny $(4)$}}\:=\:
     \frac{\hat{\sigma}_{\mbox{\tiny h}}^{7-p}}{4(6-k)(4-k)^3
     \left(1-\hat{\sigma}_{\mbox{\tiny h}}^{2(7-p)}\right)^3}
     \sum_{i=0}^{5}\mathtt{b}_{\mbox{\tiny $i$}}^{\mbox{\tiny $(p,k)$}}\hat{\sigma}_{\mbox{\tiny h}}^{i(7-p)},
    \\
   &y_{\mbox{\tiny $0$}}^{\mbox{\tiny $(6)$}}\:=\:
    -\frac{\hat{\sigma}_{\mbox{\tiny h}}^{7-p}}{24(8-k)(6-k)(4-k)^5
     \left(1-\hat{\sigma}_{\mbox{\tiny h}}^{2(7-p)}\right)^5}
     \sum_{i=0}^{9}\mathtt{c}_{\mbox{\tiny $i$}}^{\mbox{\tiny $(p,k)$}}\hat{\sigma}_{\mbox{\tiny h}}^{i(7-p)}
 \end{split}
\end{equation}
where the coefficients $\mathtt{b}_{\mbox{\tiny $i$}}^{\mbox{\tiny $(p,k)$}}$ and
$\mathtt{c}_{\mbox{\tiny $i$}}^{\mbox{\tiny $(p,k)$}}$ are constants dependent on 
the spatial dimensions of the background branes $p$ and the codimensionality of the defect $k$
and are explicitly given in Appendix \ref{App1}.
At the next order in $c_{\mbox{\tiny $f$}}$, the solution in a neighbourhood of $\varrho\sim0$ can be
generally written as 
\begin{equation}\eqlabel{y1r}
 y_{\mbox{\tiny $1$}}\left(\varrho\right)\:=\:
  \frac{\mathfrak{a}_{\mbox{\tiny $-(2-k)$}}}{\varrho^{2-k}}+
  \mathfrak{a}_{\mbox{\tiny $0$}}+\mathfrak{b}_{\mbox{\tiny $0$}}\log{\varrho}+
  \mathfrak{a}_{\mbox{\tiny $(2-k)$}}\varrho^{2-k}+\mathcal{O}\left(\varrho^{2(2-k)}\right).
\end{equation}
The coefficient of the leading order 
$\mathfrak{a}_{\mbox{\tiny $-(2-k)$}}$ and the zero order coefficient $\mathfrak{a}_{\mbox{\tiny $0$}}$
are constant of integration which will be fixed by matching the solutions of region 1 and 2.
For $k\,=\,0$ only the coefficients related to even powers of $\varrho$ are non-vanishing, while for
$k=1$ both even and odd powers are admitted. A crucial difference between the cases $k=0$ and
$k=1$ is the presence of logarithmic terms in the $k=0$ expansion, which do not instead appear for 
$k=1$. The coefficient of the logarithmic term $\mathfrak{b}_{\mbox{\tiny $0$}}$ for $k=0$ is fixed
in terms of the integration constant $\mathfrak{a}_{\mbox{\tiny $-2$}}$
\begin{equation}\eqlabel{b0log}
 \mathfrak{b}_{\mbox{\tiny $0$}}\:=\:\hat{\sigma}_{\mbox{\tiny h}}^{7-p}
  \frac{4(6-p)(5-p)-(p^2+60-29)\hat{\sigma}_{\mbox{\tiny h}}^{7-p}
        +4(7-p)(5-p)\hat{\sigma}_{\mbox{\tiny h}}^{2(7-p)}+
         4\hat{\sigma}_{\mbox{\tiny h}}^{3(7-p)}}{2\left(1-\hat{\sigma}_{\mbox{\tiny h}}^{2(7-p)}\right)^2}
        \mathfrak{a}_{\mbox{\tiny $-2$}}.
\end{equation}
In the expansion for $k=1$, the coefficient $\mathfrak{a}_{\mbox{\tiny $1$}}$ is determined in terms of
the integration constant $\mathfrak{a}_{\mbox{\tiny $-1$}}$:
\begin{equation}\eqlabel{am1}
 \mathfrak{a}_{\mbox{\tiny $1$}}\:=\:\frac{(7-p)\hat{\sigma}_{\mbox{\tiny h}}^{7-p}}{2
  \left(1-\hat{\sigma}_{\mbox{\tiny h}}^{2(7-p)}\right)^2}
  \left[
   (11-2p)+6\hat{\sigma}_{\mbox{\tiny h}}^{7-p}+(11-2p)\hat{\sigma}_{\mbox{\tiny h}}^{2(7-p)}
  \right] \mathfrak{a}_{\mbox{\tiny $-1$}}.
\end{equation}
\newline

Let us now consider the region 2, by recasting the equation \eqref{eqomYF2} in terms of the
independent variable $\tau$. For the solution one can consider the following perturbative expansion
\begin{equation}\eqlabel{yexp2}
 y\left(\tau\right)\:=\:\sum_{i=0}^{\infty}
  c_{\mbox{\tiny $f$}}^{i}
  z_{\mbox{\tiny $i$}}\left(\tau\right).
\end{equation}
Also, the position of the horizon $\tau_{\mbox{\tiny h}}$ admits a perturbative expansion for small
$c_{\mbox{\tiny $f$}}$
\begin{equation}\eqlabel{thexp}
 \tau_{\mbox{\tiny h}}\:=\:\sum_{i=0}^{\infty}
  c_{\mbox{\tiny $f$}}^{i}\tau_{\mbox{\tiny $i$}}.
\end{equation}
The 0-th order equation is
\begin{equation}\eqlabel{eqz0}
 \begin{split}
  &0\:=\:\frac{\ddot{z}_{\mbox{\tiny $0$}}}{\dot{z}_{\mbox{\tiny $0$}}^2}+
        (3-k)\left[1+\left(\frac{\sigma_{\mbox{\tiny h}}}{z_{\mbox{\tiny $0$}}}\right)^{7-p}
         \right]^{4\frac{3-k}{7-p}}\tau^{5-2k}\dot{z}_{\mbox{\tiny $0$}}+\\
  &\phantom{0\:=\:}
      +2\left(\frac{\hat{\sigma}_{\mbox{\tiny h}}}{z_{\mbox{\tiny $0$}}}\right)^{7-p}
        \frac{\tau\dot{z}_{\mbox{\tiny $0$}}-z_{\mbox{\tiny $0$}}}{z_{\mbox{\tiny $0$}}^2
         \left[1-\left(\frac{\hat{\sigma}_{\mbox{\tiny h}}}{z_{\mbox{\tiny $0$}}}
           \right)^{2\left(7-p\right)}\right]}
        \left[6-p+\left(\frac{\sigma_{\mbox{\tiny h}}}{z_{\mbox{\tiny $0$}}}\right)^{7-p}\right],
 \end{split}
\end{equation}
where the dot $\dot{\phantom{z}}$ indicates the derivative with respect to $\tau$. The conditions
at the horizon and at the boundary are
\begin{equation}\eqlabel{bcz}
 \begin{split}
  &\mbox{horizon: }\quad
   z_{\mbox{\tiny $0$}}\left(\tau_{\mbox{\tiny $0$}}\right)\:=\:\sigma_{\mbox{\tiny h}},\qquad
   z_{\mbox{\tiny $0$}}'\:=\:\frac{\sigma_{\mbox{\tiny h}}}{\tau_{\mbox{\tiny $0$}}},\\
  &\mbox{boundary: }
   z_{\mbox{\tiny $0$}}\left(\tau_{\mbox{\tiny $0$}}\right)\:=\:1
 \end{split}
\end{equation}
In order to match the solutions in region 1 and 2, it is suitable to look in the region 2 for a solution 
at large $\tau$
\begin{equation}\eqlabel{z0exp}
 z_{\mbox{\tiny $0$}}\left(\tau\right)\:=\:
   1+\sum_{i=1}^{\infty}\frac{\zeta_{\mbox{\tiny $(2-k)i$}}}{\tau^{(2-k)i}},
\end{equation}
where the very first coefficient $\zeta_{\mbox{\tiny $(2-k)$}}$ is
\begin{equation}\eqlabel{z0coeffs}
 \begin{split}
  &\zeta_{\mbox{\tiny $(2-k)$}}\:=\:-\frac{1}{(2-k)
    \left(1+\hat{\sigma}_{\mbox{\tiny h}}^{7-p}\right)^{2\frac{3-k}{7-p}}}\\
 \end{split}
\end{equation}
At first order, the perturbative solution shows a singular term, which is of order 
$\mathcal{O}\left(\tau\right)$:
\begin{equation}\eqlabel{z1exp}
 z_{\mbox{\tiny $1$}}\left(\tau\right)\:=\:
   \mathfrak{v}_{\mbox{\tiny $2-k$}}\tau^{2-k}+\mathfrak{v}_{\mbox{\tiny $0$}}+
   \mathfrak{w}_{\mbox{\tiny $0$}}\log{\tau}+
   \frac{\mathfrak{v}_{\mbox{\tiny $-(2-k)$}}}{\tau^{2-k}}+\mathcal{O}\left(\frac{1}{\tau^{2(2-k)}}\right),
\end{equation}
where $\mathfrak{v}_{\mbox{\tiny $0$}}$ is an integration constant and the coefficients 
$\mathfrak{v}_{\mbox{\tiny $2-k$}}$ and $\mathfrak{w}_{\mbox{\tiny $0$}}$ are fixed in terms of 
the zero-th order solution to be:
\begin{equation}\eqlabel{vcoeffs}
 \mathfrak{v}_{\mbox{\tiny $2-k$}}\:=\: 
     \left\{
      \begin{array}{l}
       \left.\frac{(3+k-p)+(4-k)\hat{\sigma}_{\mbox{\tiny h}}^{7-p}}{(4-k)
        \left(1-\hat{\sigma}_{\mbox{\tiny h}}^{2(7-p)}\right)}\,\hat{\sigma}_{\mbox{\tiny h}}^{7-p}
       \right|_{\mbox{\tiny $k=0$}},\qquad\mbox{ for $k=0$}\\
       0,\hspace{4.5cm}\qquad\mbox{ for $k=1$}.
      \end{array}
     \right.
\end{equation}
We are now in condition to compare the solutions in the two different regions of the radial coordinate.
First, it is convenient to rewrite the expansion in the region 2 just in terms of the radial coordinate
$\varrho$ as well as the density $c_{\mbox{\tiny $f$}}$. We will rewrite the expansions both in region 1
and 2 so that the comparison between the two results becomes straightforward
\begin{equation}\eqlabel{zexp}
 \begin{split}
  &\left.y\left(\rho\right)\right|_{\mbox{\tiny reg 1}}\:=\:
   1+y_{\mbox{\tiny $0$}}^{\mbox{\tiny (2)}}\varrho^2+\mathcal{O}\left(\varrho^4\right)+
   c_{\mbox{\tiny $f$}}
   \left[
    \frac{\mathfrak{a}_{\mbox{\tiny $-(2-k)$}}}{\varrho^{2-k}}+
  \mathfrak{a}_{\mbox{\tiny $0$}}+\mathfrak{b}_{\mbox{\tiny $0$}}\log{\varrho}+
  \mathfrak{a}_{\mbox{\tiny $(2-k)$}}\varrho^{2-k}+\mathcal{O}\left(\varrho^{2(2-k)}\right)
   \right],\\
  &\left.y\left(\rho\right)\right|_{\mbox{\tiny reg 2}}\:=\:
   1+\mathfrak{v}_{\mbox{\tiny $(2-k)$}}\varrho^{2-k}+\mathfrak{y}_{\mbox{\tiny $2(2-k)$}}\varrho^{2(2-k)}+
    \mathcal{O}\left(\varrho^{3(2-k)}\right)+\\
  &\phantom{\left.y\left(\rho\right)\right|_{\mbox{\tiny reg 2}}\:=\:1 }
   + c_{\mbox{\tiny $f$}}
   \left[
    \frac{\zeta_{2-k}}{\varrho^{2-k}}+\mathfrak{v}_{\mbox{\tiny $0$}}+
     \mathfrak{w}_{\mbox{\tiny $0$}}\log{\varrho}-
     \frac{\mathfrak{w}_{\mbox{\tiny $0$}}}{2-k}\log{c_{\mbox{\tiny $f$}}}
     +\mathcal{O}\left(\varrho^{2-k}\right)
   \right],
 \end{split}
\end{equation}
where the coefficient $\mathfrak{y}_{\mbox{\tiny $2(2-k)$}}$ comes from the equations at order 
$c_{\mbox{\tiny $f$}}^2$, which we have not written down explicitly. The matching at zero-th order in 
$c_{\mbox{\tiny $f$}}$ fixes some of the integration constants and provides some non-trivial consistency 
check. Notice that this matching provides a non-trivial check on the expansion: 
\begin{equation}\eqlabel{match1}
 \begin{split}
  &\left.\mathfrak{v}_{2-k}\right|_{\mbox{\tiny k=0}}\:=\:y_{\mbox{\tiny $0$}}^{\mbox{\tiny (2)}}\\
  &\left.\mathfrak{v}_{2-k}\right|_{\mbox{\tiny k=1}}\:=\:0,
 \end{split}
\end{equation}
where the coefficients in \eqref{match1} are provided in \eqref{y0rCoeffs} and \eqref{vcoeffs} and 
identically satisfy the equalities in \eqref{match1}.
The matching at first order in $c_{\mbox{\tiny $f$}}$ fixes some integration constants:
\begin{equation}\eqlabel{match2}
 \mathfrak{a}_{\mbox{\tiny $-(2-k)$}}\:=\:\zeta_{2-k},
 \qquad
 \mathfrak{w}_{\mbox{\tiny $0$}}\:=\:\mathfrak{b}_{\mbox{\tiny $0$}},
 \qquad
 \mathfrak{v}_{\mbox{\tiny $0$}}\:=\:\mathfrak{a}_{\mbox{\tiny $0$}}+
  \frac{\mathfrak{w}_{\mbox{\tiny $0$}}}{2-k}\log{c_{\mbox{\tiny $f$}}},
\end{equation}
where the second equality becomes the identity $0=0$ for $k=1$. The constant of integration 
$\mathfrak{a}_{\mbox{\tiny $0$}}$ is determined by the boundary condition 
$\left.y_{\mbox{\tiny $1$}}\left(\varrho\right)\right|_{\mbox{\tiny $\varrho\rightarrow\infty$}}\rightarrow0$.

This small density analysis of the embedding function showed that the appearance of logarithmic terms
is common to all the systems with no defect while it disappears when a codimension-1 defect is introduced.

\subsubsection{Small density expansion of chemical potential}

We can now use the previous analysis to explicitly compute the chemical potential in the small density limit.
The chemical potential is given by the integral in \eqref{cp} with extreme of integration 
$[\varrho_{\mbox{\tiny $\Lambda$}},\,\infty[$ and 
$[\varrho_{\mbox{\tiny h}},\,\varrho_{\mbox{\tiny $\Lambda$}}]$ respectively in region 1 and region 2.

In region 1, the leading term of the chemical potential in a $c_{\mbox{\tiny $f$}}$ perturbative expansion is
of order one
\begin{equation}\eqlabel{cpout}
 \left.\mu\right|_{\mbox{\tiny reg 1}}\:=\:
  c_{\mbox{\tiny $f$}}
  \int_{\varrho_{\mbox{\tiny $\Lambda$}}}^{\infty}d\varrho\:
  \left(\varrho^2+y_{\mbox{\tiny $0$}}^2\right)^{2-k}
  \frac{\left(\varrho^2+y_{\mbox{\tiny $0$}}^2\right)^{\frac{7-p}{2}}-\hat{\sigma}_{\mbox{\tiny h}}^{7-p}}{
   \left[
    \left(\varrho^2+y_{\mbox{\tiny $0$}}^2\right)^{\frac{7-p}{2}}+\hat{\sigma}_{\mbox{\tiny h}}^{7-p}
   \right]^{\frac{11-p-2k}{7-p}}}
  \frac{\sqrt{1+\left(y_{\mbox{\tiny $0$}}'\right)^2}}{\varrho^{3-k}}+
  \mathcal{O}\left(c_{\mbox{\tiny $f$}}^2\right).
\end{equation}
Using the perturbative solution \eqref{y0r} for $y_{\mbox{\tiny $0$}}$, it is easy to compute the first 
orders of the chemical potential in a neighbourhood of $\varrho\,\sim\,0$
\begin{equation}\eqlabel{cpreg1}
 \left.\mu^{\mbox{\tiny $(1)$}}\right|_{\mbox{\tiny reg 1}}\:=\:
  \frac{\mathfrak{m}_{\mbox{\tiny $-(2-k)$}}}{\varrho_{\mbox{\tiny $\Lambda$}}^{2-k}}+
  \mathfrak{m}_{\mbox{\tiny $0$}}\log{\varrho_{\mbox{\tiny $\Lambda$}}}+
  \mathfrak{m}_{\mbox{\tiny $1$}}\varrho_{\mbox{\tiny $\Lambda$}}+
  \mathcal{K}\left(\varrho_{\mbox{\tiny $\Lambda$}}\right),
\end{equation}
where
\begin{equation}\eqlabel{ms}
 \begin{split}
  &\mathfrak{m}_{\mbox{\tiny $-(2-k)$}}\:=\:\frac{1-\hat{\sigma}_{\mbox{\tiny h}}^{7-p}}{(2-k)
    \left(1+\hat{\sigma}_{\mbox{\tiny h}}^{7-p}\right)^{\frac{11-p-2k}{7-p}}},\\
  &\mathfrak{m}_{\mbox{\tiny $0$}}\:=\:\delta_{\mbox{\tiny $k,0$}}\mathfrak{m}_{\mbox{\tiny $k$}},
   \hspace{4cm}
   \mathfrak{m}_{\mbox{\tiny $1$}}\:=\:\delta_{\mbox{\tiny $k,1$}}\mathfrak{m}_{\mbox{\tiny $k$}},\\
  &\mathfrak{m}_{\mbox{\tiny $k$}}\:=\:
    \frac{1}{(4-k)^2\left(1-\hat{\sigma}_{\mbox{\tiny h}}^{2(7-p)}\right)
     \left(1+\hat{\sigma}_{\mbox{\tiny h}}^{(7-p)}\right)^{2\frac{9-p-k}{7-p}}}
    \sum_{i=0}^{4}\mathfrak{c}_{i}^{\mbox{\tiny $(p,k)$}}\hat{\sigma}_{\mbox{\tiny h}}^{i(7-p)},
   \end{split}
\end{equation}
and the last term $\mathcal{K}\left(\varrho_{\mbox{\tiny $\Lambda$}}\right)$ in 
\eqref{cpreg1} is finite and defined by
\begin{equation}\eqlabel{Kfin}
 \mathcal{K}\left(\varrho_{\mbox{\tiny $\Lambda$}}\right)\:=\:
    \int_{\varrho_{\mbox{\tiny $\Lambda$}}}^{\infty}d\varrho\:
    \left.\frac{f'\left(\varrho\right)}{c_{\mbox{\tiny $f$}}}\right|_{
     \mbox{\tiny $c_{\mbox{\tiny $f$}}=0$}}^{
     \mbox{\tiny $y=y_{\mbox{\tiny $0$}}$}}
   -\frac{\mathfrak{m}_{\mbox{\tiny $-(2-k)$}}}{\varrho_{\mbox{\tiny $\Lambda$}}^{2-k}}-
   \mathfrak{m}_{\mbox{\tiny $0$}}\log{\varrho_{\mbox{\tiny $\Lambda$}}}-
   \mathfrak{m}_{\mbox{\tiny $1$}}\varrho_{\mbox{\tiny $\Lambda$}}
\end{equation}

In region 2, the small density expansion shows both a zero-th and first order terms
\begin{equation}\eqlabel{cpin}
 \left.\mu\right|_{\mbox{\tiny reg 2}}\:=\:
  \int_{\tau_{\mbox{\tiny $h$}}}^{\Lambda}d\tau\:
  \frac{\mathtt{h}_{\mbox{\tiny $-$}}}{\mathtt{h}_{\mbox{\tiny $+$}}^{\frac{5-p}{7-p}}}
  \frac{\sqrt{c_{\mbox{\tiny $f$}}^{\frac{2}{2-k}}+\dot{y}^2}}{
   \sqrt{1+c_{\mbox{\tiny $f$}}^{\frac{2}{2-k}}\mathtt{h}_{\mbox{\tiny $+$}}^{4\frac{3-k}{7-p}}}}
  \:=\:
  \left.\mu^{\mbox{\tiny $(0)$}}\right|_{\mbox{\tiny reg 2}}+
  c_{\mbox{\tiny $f$}}\left.\mu^{\mbox{\tiny $(1)$}}\right|_{\mbox{\tiny reg 2}}+
  \mathcal{O}\left( c_{\mbox{\tiny $f$}}^2\right),
\end{equation}
where the zero-th order term, which is fixed by the solution \eqref{z0exp} for $z_{\mbox{\tiny $0$}}$,
can be conveniently written as
\begin{equation}\eqlabel{cp0in}
 \left.\mu^{\mbox{\tiny $(0)$}}\right|_{\mbox{\tiny reg 2}}\:=\:
  m-
  \int_{\Lambda}^{\infty}d\tau\:\frac{\dot{z}_{\mbox{\tiny $0$}}}{z_{\mbox{\tiny $0$}}^2}
   \frac{z_{\mbox{\tiny $0$}}^{7-p}-\hat{\sigma}_{\mbox{\tiny h}}^{7-p}}{
    \left(z_{\mbox{\tiny $0$}}^{7-p}+\hat{\sigma}_{\mbox{\tiny h}}^{7-p}\right)^{\frac{5-p}{7-p}}}.
\end{equation}
Using the perturbative expansion \eqref{z1exp} as $\tau\,\rightarrow\,\infty$, the zero-th order
chemical potential can be obtained as an expansion in $\Lambda^{-1}$
\begin{equation}\eqlabel{cp0in2}
 \left.\mu^{\mbox{\tiny $(0)$}}\right|_{\mbox{\tiny reg 2}}\:=\:
  m-\frac{1-\hat{\sigma}_{\mbox{\tiny h}}^{7-p}}{(2-k)
   \left(1+\hat{\sigma}_{\mbox{\tiny h}}^{7-p}\right)^{\frac{11-p-2k}{7-p}}}
   \frac{1}{\Lambda^{2-k}}+\mathcal{O}\left(\frac{1}{\Lambda^{2(2-k)}}\right).
\end{equation}
At first order in $c_{\mbox{\tiny $f$}}$, the chemical potential receives contributions from
the zero-th order solution $z_{\mbox{\tiny $0$}}$ in \eqref{z0exp} as well as the first order
one $z_{\mbox{\tiny $1$}}$ in \eqref{z1exp}. In full generality, it can be written as
\begin{equation}\eqlabel{cp1in}
 \begin{split}
  &\left.\mu^{\mbox{\tiny $(1)$}}\right|_{\mbox{\tiny reg 2}}\:=\:
   \int_{\tau_{\mbox{\tiny $0$}}}^{\Lambda}d\tau\:
   \left\{
    \partial_{\mbox{\tiny $\tau$}}
    \left[
     \frac{z_{\mbox{\tiny $0$}}^{7-p}-\hat{\sigma}_{\mbox{\tiny h}}^{7-p}}{z_{\mbox{\tiny $0$}}^2
      \left(z_{\mbox{\tiny $0$}}^{7-p}+\sigma_{\mbox{\tiny h}}^{7-p}\right)^{\frac{5-p}{7-p}}}
     z_{\mbox{\tiny $1$}}
    \right]
   +\delta_{k,0}\left[
     \frac{
      z_{\mbox{\tiny $0$}}^{7-p}-\hat{\sigma}_{\mbox{\tiny h}}^{7-p}}{
      2z_{\mbox{\tiny $0$}}^{2}\dot{z}_{\mbox{\tiny $0$}}
     \left(z_{\mbox{\tiny $0$}}^{7-p}+\hat{\sigma}_{\mbox{\tiny h}}^{7-p}\right)^{\frac{5-p}{7-p}}}
    \right.\right.+\\
   &+\left.\left.
      \tau^2\frac{\left[2z_{\mbox{\tiny $0$}}^{10}\hat{\sigma}_{\mbox{\tiny h}}^{7-p}
      \left(\hat{\sigma}_{\mbox{\tiny h}}^{7-p}+(6-p)z_{\mbox{\tiny $0$}}^{7-p}\right)-
      \tau^4\left(z_{\mbox{\tiny $0$}}^{2(7-p)}-\hat{\sigma}_{\mbox{\tiny h}}^{2(7-p)}\right)
      \left(z_{\mbox{\tiny $0$}}^{7-p}+\hat{\sigma}_{\mbox{\tiny h}}^{7-p}\right)^{\frac{12}{7-p}}
      \right]\dot{z}_{\mbox{\tiny $0$}}}{2z_{\mbox{\tiny $0$}}^{14}
     \left(z_{\mbox{\tiny $0$}}^{7-p}+\hat{\sigma}_{\mbox{\tiny h}}^{7-p}\right)^{2\frac{6-p}{7-p}}}
    \right]
   \right\}
 \end{split}
\end{equation}
The first term in \eqref{cp1in} can be easily integrated. Notice that in the case of codimension-$1$ systems,
it is the only term contributing to $\left.\mu^{\mbox{\tiny $\left(1\right)$}}\right|_{\mbox{\tiny reg 2}}$.
Using the expansions \eqref{z0exp} and \eqref{z1exp}, as well as the relations \eqref{match1} and 
\eqref{match2}, in the case of $k=1$ it acquires the form
\begin{equation}\eqlabel{m1k1}
 \left.\mu^{\mbox{\tiny $(1)$}}\right|_{\mbox{\tiny reg 2}}^{\mbox{\tiny $k=1$}}\:=\:
  \frac{1-\hat{\sigma}^{7-p}}{\left(1+\hat{\sigma}^{7-p}\right)^{\frac{5-p}{7-p}}}
   \mathfrak{v}_{\mbox{\tiny $0$}}.
\end{equation}
Therefore one can write down the chemical potential in the small density expansion as follows
\begin{equation}\eqlabel{mTfink1}
 \mu\:=\:m+\tilde{\mathfrak{s}}\left(T\right)c_{\mbox{\tiny $f$}}
          +\mathcal{O}\left(c_{\mbox{\tiny $f$}}^2\right),
\end{equation}
where $\tilde{\mathfrak{s}}\left(T\right)$ is defined as
\begin{equation}\eqlabel{s}
 \tilde{\mathfrak{s}}\left(T\right)\:=\:\lim_{\varrho_{\mbox{\tiny $\Lambda$}}\rightarrow0}
  \mathcal{K}\left(\varrho_{\mbox{\tiny $\Lambda$}}\right)+
  \left.\mu^{\mbox{\tiny $(1)$}}\right|_{\mbox{\tiny reg 2}},
\end{equation}
which should vanish in the zero temperature limit.
In principle one would need to fix the constant $\mathfrak{v}_{\mbox{\tiny $0$}}$ by imposing the boundary
condition $\left.y_{\mbox{\tiny $1$}}\right|_{\mbox{\tiny $\varrho\rightarrow\infty$}}\,=\,0$. However,
for our purposes, {\it i.e.} to check the order of the phase transition, it will not be necessary. 
The function $\tilde{\mathfrak{s}}\left(T\right)$ in \eqref{mTfink1} may be both non-zero and zero.
In the second case one would need to go to next order in $c_{\mbox{\tiny $f$}}$, in which case the small
density expansion of the chemical potential $\mu$ can {\it at most} acquire the form:
\begin{equation}\eqlabel{mTfink1b}
 \mu\:=\:m+\mathfrak{s}_{\mbox{\tiny $1$}}\left(T\right)c_{\mbox{\tiny $f$}}^2-
           \mathfrak{s}_{\mbox{\tiny $2$}}\left(T\right)c_{\mbox{\tiny $f$}}^2
            \log{c_{\mbox{\tiny $f$}}}
\end{equation}
However, in any case the order of the phase transition will be the same. We will comment on this in the next 
subsection. For the time being, let us focus now on the structure of the small density expansion of the 
chemical potential for system with no defect.


In the case of $k=0$ we have a non-trivial contribution to 
$\left.\mu^{\mbox{\tiny $(1)$}}\right|_{\mbox{\tiny reg 2}}^{\mbox{\tiny $k=0$}}$
from both the terms in \eqref{cp1in}. For convenience let us write \eqref{cp1in} as
\begin{equation}\eqlabel{cp1inB}
 \left.\mu^{\mbox{\tiny $(1)$}}\right|_{\mbox{\tiny reg 2}}^{\mbox{\tiny $k=0$}}\:=\:
  \left.\mu^{\mbox{\tiny $(1)$}}_{\mbox{\tiny a}}\right|_{\mbox{\tiny reg 2}}+
  \left.\mu^{\mbox{\tiny $(1)$}}_{\mbox{\tiny b}}\right|_{\mbox{\tiny reg 2}},
\end{equation}
where $\left.\mu^{\mbox{\tiny $(1)$}}_{\mbox{\tiny a}}\right|_{\mbox{\tiny reg 2}}$ indicates the total term in
\eqref{cp1in}. The first term in \eqref{cp1inB} is easy to obtain and can be written as
\begin{equation}\eqlabel{cp1inBa}
 \begin{split}
 \left.\mu^{\mbox{\tiny $(1)$}}_{\mbox{\tiny a}}\right|_{\mbox{\tiny reg 2}}\:&=\:
  \frac{3-p+4\hat{\sigma}_{\mbox{\tiny h}}^{7-p}}{
   4\left(1+\hat{\sigma}_{\mbox{\tiny h}}^{7-p}\right)^{2\frac{6-p}{7-p}}}\hat{\sigma}_{\mbox{\tiny h}}^{7-p}
   \Lambda^{2}+
   \frac{1-\hat{\sigma}_{\mbox{\tiny h}}^{7-p}}{\left(1+\hat{\sigma}_{\mbox{\tiny h}}^{7-p}\right)^{\frac{5-p}{7-p}}}
    \mathfrak{v}_{\mbox{\tiny $0$}}-\\
   &\phantom{=\:}-
   \frac{(6-p)(3-p)+(27-5p)\hat{\sigma}_{\mbox{\tiny h}}^{7-p}+4\hat{\sigma}_{\mbox{\tiny h}}^{2(7-p)}}{4
    \left(1-\hat{\sigma}_{\mbox{\tiny h}}^{2(7-p)}\right)\left(1+\hat{\sigma}_{\mbox{\tiny h}}^{7-p}\right)^{
     2\frac{9-p}{7-p}}}\hat{\sigma}_{\mbox{\tiny h}}^{2(7-p)}+\\
 &\phantom{=\:}+
  \frac{4(6-p)(3-p)\left(1-\hat{\sigma}_{\mbox{\tiny h}}^{2(7-p)}\right)+(199-26p-p^2)\hat{\sigma}_{\mbox{\tiny h}}^{7-p}}{
   8\left(1-\hat{\sigma}_{\mbox{\tiny h}}^{2(7-p)}\right)
   \left(1+\hat{\sigma}_{\mbox{\tiny h}}^{7-p}\right)^{2\frac{9-p}{7-p}}}\log{\Lambda},
 \end{split}
\end{equation}
where, similarly to \eqref{m1k1}, we would not need to fix the coefficient $\mathfrak{v}_{\mbox{\tiny $0$}}$ explicitly.
The second term in \eqref{cp1inB} instead acquires the form
\begin{equation}\eqlabel{cp1inBb}
 \begin{split}
  \left.\mu^{\mbox{\tiny $(1)$}}_{\mbox{\tiny b}}\right|_{\mbox{\tiny reg 2}}\:&=\:
   -\frac{3-p+4\hat{\sigma}_{\mbox{\tiny h}}^{7-p}}{
   4\left(1+\hat{\sigma}_{\mbox{\tiny h}}^{7-p}\right)^{2\frac{6-p}{7-p}}}\hat{\sigma}_{\mbox{\tiny h}}^{7-p}
   \Lambda^{2}-\\
  &-\left\{
    \frac{2(p^2-9p+12)+(53+14p-3p^2)\hat{\sigma}_{\mbox{\tiny h}}^{7-p}-2(12+5p-p^2)\hat{\sigma}_{\mbox{\tiny h}}^{2(7-p)}-
     4\hat{\sigma}_{\mbox{\tiny h}}^{3(7-p)}}{4\left(1+\hat{\sigma}_{\mbox{\tiny h}}^{7-p}\right)^{\frac{25-3p}{7-p}}
     \left(1-\hat{\sigma}_{\mbox{\tiny h}}^{7-p}\right)}-\right.\\
  &\phantom{\{}\left.-
     \frac{6-p+(7-p)\hat{\sigma}_{\mbox{\tiny h}}^{7-p}+\hat{\sigma}_{\mbox{\tiny h}}^{2(7-p)}}{
     \left(1+\hat{\sigma}_{\mbox{\tiny h}}^{7-p}\right)^{\frac{25-3p}{7-p}}}
   \right\}\hat{\sigma}_{\mbox{\tiny h}}^{7-p}\log{\Lambda}+\mathcal{W}
 \end{split}
\end{equation}
where, similarly to $\mathcal{K}$, $\mathcal{W}$, is defined as
\begin{equation}\eqlabel{Hfin}
 \mathcal{W}=\:\lim_{\mbox{\tiny $\Lambda\rightarrow\infty$}}
  \left.\mu^{\mbox{\tiny $(1)$}}_{\mbox{\tiny b}}\right|_{\mbox{\tiny reg 2}}
  +\frac{3-p+4\hat{\sigma}_{\mbox{\tiny h}}^{7-p}}{
   4\left(1+\hat{\sigma}_{\mbox{\tiny h}}^{7-p}\right)^{2\frac{6-p}{7-p}}}\hat{\sigma}_{\mbox{\tiny h}}^{7-p}
   \Lambda^{2}+\mathcal{C}\left(\hat{\sigma}_{\mbox{\tiny h}}^{7-p}\right)\log{\Lambda},
\end{equation}
where $\mathcal{C}\left(\hat{\sigma}_{\mbox{\tiny h}}^{7-p}\right)$ indicates the term of \eqref{cp1inBb} in curl bracket.
Notice that the first (divergent) term in \eqref{cp1inBb} exactly cancels the quadratic divergence (as 
$\Lambda\rightarrow\infty$) in \eqref{cp1inBa}.
Summing the contributions \eqref{cpreg1}, \eqref{cp0in2} ,\eqref{cp1inBa} and \eqref{cp1inBb}, as well as using the third 
relation \eqref{match2}, the chemical potential in the small density expansion can be expressed as
\begin{equation}\eqlabel{mTfink0}
 \mu\:=\:m+\mathfrak{s}_{\mbox{\tiny $1$}}\left(T\right)c_{\mbox{\tiny $f$}}-
           \mathfrak{s}_{\mbox{\tiny $2$}}\left(T\right)c_{\mbox{\tiny $f$}}
            \log{c_{\mbox{\tiny $f$}}},
\end{equation}
where $\mathfrak{s}_{\mbox{\tiny $1$}}\left(T\right)$ and $\mathfrak{s}_{\mbox{\tiny $2$}}\left(T\right)$ 
are given respectively by
\begin{equation}\eqlabel{s1s2k0}
 \begin{split}
 &\mathfrak{s}_{\mbox{\tiny $1$}}\left(T\right)\:=\:\mathcal{K}+
  \frac{1-\hat{\sigma}_{\mbox{\tiny h}}^{7-p}}{\left(1+\hat{\sigma}_{\mbox{\tiny h}}^{7-p}\right)^{\frac{5-p}{7-p}}}
  \mathfrak{a}_{\mbox{\tiny $0$}}
 -\frac{(6-p)(3-p)+(27-5p)\hat{\sigma}_{\mbox{\tiny h}}^{7-p}+4\hat{\sigma}_{\mbox{\tiny h}}^{2(7-p)}}{4
    \left(1-\hat{\sigma}_{\mbox{\tiny h}}^{2(7-p)}\right)\left(1+\hat{\sigma}_{\mbox{\tiny h}}^{7-p}\right)^{
     2\frac{9-p}{7-p}}}\hat{\sigma}_{\mbox{\tiny h}}^{2(7-p)}+\mathcal{W}\\
 &\mathfrak{s}_{\mbox{\tiny $2$}}\left(T\right)\:=\:
   \frac{1}{2}\left(\mathtt{x}-\mathtt{y}-\mathfrak{b}_{\mbox{\tiny $0$}}\right),
\end{split}
\end{equation}
where $\mathtt{x}$ and $\mathtt{y}$ are the coefficients of $\log{\Lambda}$ in \eqref{cp1inBa} and 
\eqref{cp1inBb} respectively, and $\mathfrak{b}_{\mbox{\tiny $0$}}$ is given in \eqref{b0log}.
The expressions \eqref{mTfink1} and \eqref{mTfink0} can be written as
\begin{equation}\eqlabel{mTfink}
\mu\:=\:m+\mathfrak{s}_{\mbox{\tiny $1$}}\left(T\right)c_{\mbox{\tiny $f$}}^{\frac{2}{2-k}}-
           \mathfrak{s}_{\mbox{\tiny $2$}}\left(T\right)c_{\mbox{\tiny $f$}}^{\frac{2}{2-k}}
            \log{c_{\mbox{\tiny $f$}}}
\end{equation}
In the next subsection, we will discuss in detail the order the phase transition in the interior of the
plane $(\mu,\,T)$ using the results of this subsection.

\subsection{Phase transitions in the $\left(\mu,\,T\right)$-plane}

In the previous section we were able to obtain a very general {\it analytic} expression for the
chemical potential in the small density limit. Let us emphasise again an important point. The charge
density $n$, which is given in \eqref{n}, is proportional to the density $c_{\mbox{\tiny $f$}}$. The
Minkowski and black-hole embeddings are characterised by zero and non-zero $n$ respectively. Therefore, working
in the small density limit with black-hole embedding boundary conditions means focusing on a neighbourhood of 
the transition curve in the $\left(\mu,\,T\right)$-plane, so that the limit $c_{\mbox{\tiny $f$}}\rightarrow0$
send us on the transition curve.

Let us now look in more detail at this phase transition. Since we are in the grand-canonical ensemble, the
right thermodynamical potential to analyse is the grand potential $\Omega$ and its derivatives. In the
Minkowski embedding phase ({\it i.e.} $\mu\,<\,m$), all the derivatives of the grand-potential $\Omega$ turn
out to be zero:
\begin{equation}\eqlabel{MEO}
 \left.\frac{\partial\Omega}{\partial\mu}\right|_{\mbox{\tiny $\mu<m$}}\:\sim\:
  n\Big|_{\mbox{\tiny $\mu<m$}}\:=\:0,\qquad
 \left.\frac{\partial^{\mbox{\tiny $s$}}\Omega}{\partial\mu^{\mbox{\tiny $s$}}}\right|_{\mbox{\tiny $\mu<m$}}
  \:=\:0,\qquad\forall s\ge1.
\end{equation}
In the black-hole phase, ({\it i.e.} $\mu\,>\,m$), we can use the expression \eqref{mTfink} for the
chemical potential (if $\tilde{\mathfrak{s}}(T)\,=\,0$), then take the limit 
$c_{\mbox{\tiny $f$}}\rightarrow0$ to go on the transition curve $\mu\,=\,m$ and match the result with 
\eqref{MEO}:
\begin{equation}\eqlabel{BEO}
 \left.\frac{\partial^{\mbox{\tiny $2$}}\Omega}{\partial\mu^{\mbox{\tiny $2$}}}\right|_{\mbox{\tiny $\mu>m$}}
  \:\sim\:\frac{c_{\mbox{\tiny $f$}}^{-\frac{k}{2-k}}}{\frac{2}{2-k}
   \mathfrak{s}_{\mbox{\tiny $1$}}\left(T\right)-\mathfrak{s}_{\mbox{\tiny $2$}}\left(T\right)-
   \frac{2}{2-k}\mathfrak{s}_{\mbox{\tiny $2$}}\left(T\right)\log{c_{\mbox{\tiny $f$}}}}
   \:\overset{\mbox{\tiny $c_{\mbox{\tiny $f$}}\rightarrow0$}}{\longrightarrow}\:
   \left\{
    \begin{array}{l}
     0,\qquad k=0\\
     \phantom{\ldots}\\
     \infty,\qquad k=1.
    \end{array}
   \right.
\end{equation}
Already from \eqref{BEO} one can infer that, while for $k=0$ there is no discontinuity in the second 
derivative of the chemical potential, for $k=1$ it diverges. For codimension-$1$ systems, the
phase transition in the $\left(\mu,\,T\right)$-plane is of second order.

Let us now consider the third-derivative of the grand-potential $\Omega$ for $k=0$:
\begin{equation}\eqlabel{BEO2}
 \left.\frac{\partial^{\mbox{\tiny $3$}}\Omega}{\partial\mu^{\mbox{\tiny $3$}}}\right|_{
  \mbox{\tiny $\mu>m$, $k=0$}}
  \:\sim\:-
  \frac{\mathfrak{s}_{\mbox{\tiny $2$}}\left(T\right)}{c_{\mbox{\tiny $f$}}
  \left(\mathfrak{s}_{\mbox{\tiny $1$}}\left(T\right)-\mathfrak{s}_{\mbox{\tiny $2$}}\left(T\right)-
  \mathfrak{s}_{\mbox{\tiny $2$}}\left(T\right)\log{c_{\mbox{\tiny $f$}}}\right)^2}
  \:\overset{\mbox{\tiny $c_{\mbox{\tiny $f$}}\rightarrow0$}}{\longrightarrow}\:
  \infty.
\end{equation}
As a consequence of such a divergence, the phase transition for codimension-$0$ systems is of
third order.

In the case $\tilde{\mathfrak{s}}(T)\,\neq\,0$ for codimension-$1$ defects, then
the second derivative of $\Omega$ with respect the chemical potential $\mu$ acquires the form
\begin{equation}\eqlabel{BEO3}
 \left.\frac{\partial^{\mbox{\tiny $2$}}\Omega}{\partial\mu^{\mbox{\tiny $2$}}}\right|_{\mbox{\tiny $\mu>m$}}
  \:\sim\:
  \frac{1}{\tilde{\mathfrak{s}}_{1}\left(T\right)}+\mathcal{O}\left(c_{\mbox{\tiny $f$}}\right)
  \:\overset{c_{\mbox{\tiny $f$}}\rightarrow0}{\longrightarrow}\:
  \frac{1}{\tilde{\mathfrak{s}}_{1}\left(T\right)},
  \qquad k\,=\,1.
\end{equation}
Even with the form \eqref{mTfink1}, the second derivative shows a discontinuity. 
\begin{figure}
 \centering%
 {\scalebox{.6}{\includegraphics{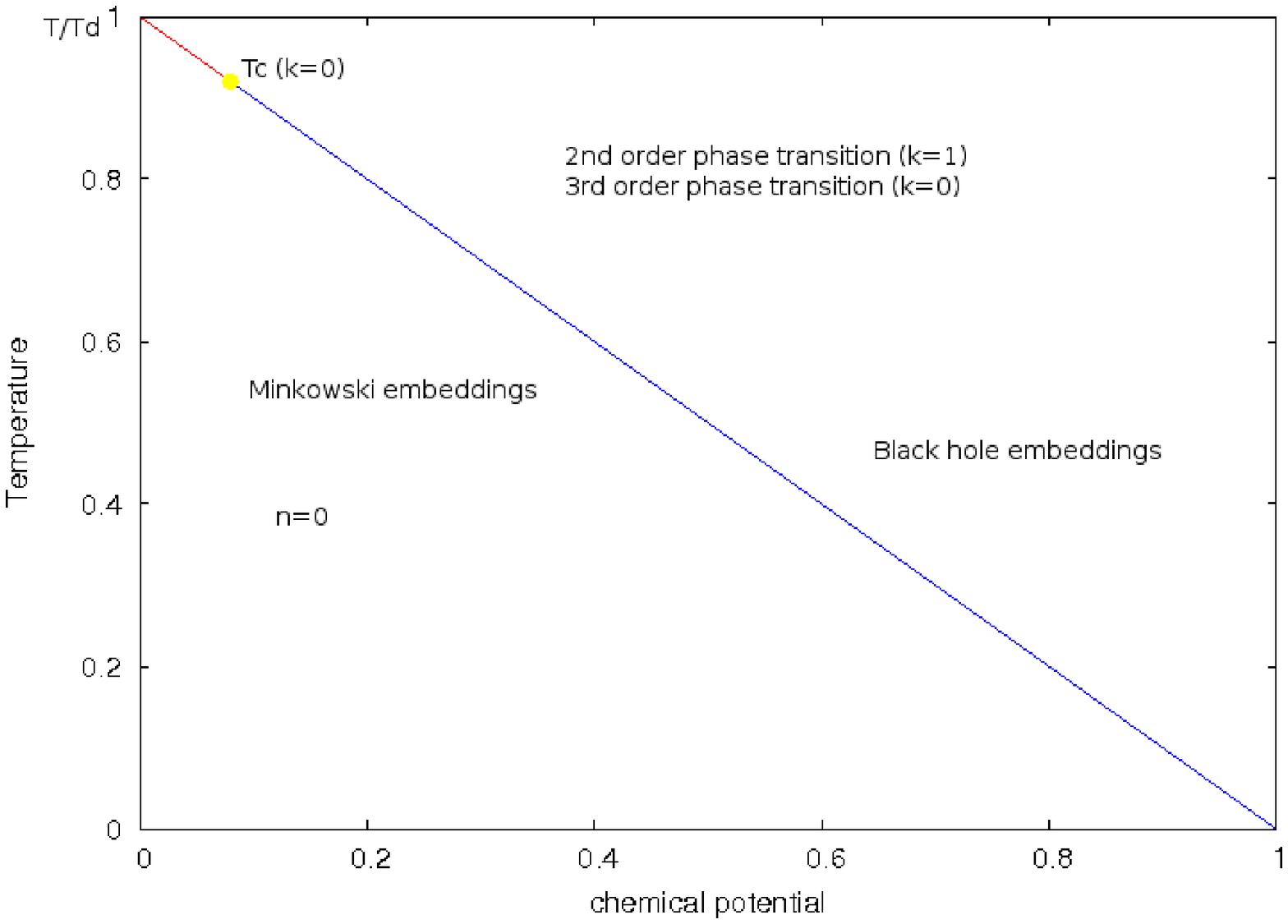}}}
 \caption{Phase diagram $(\mu,\,T)$. The axises are defined as $\mu/m$ and $T/T_{\mbox{\tiny $d$}}$. 
          The phase diagram shows a first order phase transition at $T=T_{\mbox{\tiny $d$}}$ along
          the temperature axis (systems at zero baryonic chemical potential), while at zero temperature
          and finite density all the D$p$/D$(p+4-2k)$ systems are characterised by a second order
          phase transition at $\mu\,=\,m_{\mbox{\tiny $0$}}$. The curve $\mu\,=\,m\left(T\right)$
          actually represents a transition curve in the interior of the $(\mu,\,T)$ plane. The region
          $\mu\,<\,m\left(T\right)$ corresponds to Minkowski-like embeddings with $n\,=\,0$, while
          the region $\mu\,>\,m\left(T\right)$ corresponds to a black hole phase, with the transition
          between the two phases is of second order for theories with a codimension-$1$ defect and of
          third order for theories with no defect.}
\end{figure}

One comment is now in order. The results \eqref{BEO}, \eqref{BEO2} and \eqref{BEO3} imply that the order of 
the phase transition in the $\left(\mu,\,T\right)$-plane is tied to the codimensionality of the
defect theory. We therefore identified two universality classes of theories determined by their
codimensionality, or equivalently, by the order of the phase transition in the $\left(\mu,\,T\right)$-plane.

The existence of a zero in the function $\mathfrak{s}_{\mbox{\tiny $2$}}\left(T\right)$ would implicate
the presence for $k=0$ systems of a critical point 
$\left(\mu_{\mbox{\tiny $c$}},\,T_{\mbox{\tiny $c$}}\right)$ at which the phase transition becomes of
second order.

\section{Massless hypermultiplet and quantum fluids from non-conformal branes backgrounds}\label{hql}

In order to consider massless degrees of freedom on a $(1+p-k)$-defect, we need to fix the probe branes
to wrap the maximal sphere $S^{\mbox{\tiny $3-k$}}$. The probe branes therefore wrap 
$\mathcal{M}_{\mbox{\tiny $p+2-k$}}\times S^{\mbox{\tiny $3-k$}}$. We will work directly in the 
black hole embedding: since the probe branes wrap the black hole geometry 
$\mathcal{M}_{\mbox{\tiny $p+2-k$}}$, the induced metric contains a black hole as well.
This class of embedding can therefore be described through the linear coordinate 
$x^{\mbox{\tiny $p$}}\,\equiv\,z\left(r\right)$\footnote{Here we indicate $z$ as function of $r$ since we will
consider the background line element \eqref{blackDp}.} for defect theories.  
If the embedding mode $z(r)$ has a non-trivial profile, 
its dual operator $\mathcal{O}_{\mbox{\tiny $z$}}$ acquires a non-zero vacuum-expectation-value breaking the 
supersymmetries. If the embedding mode $z(r)$ is a constant (namely $z=0$), the probe branes have a fixed 
position in the $(p+2-k)$-dimensional non-compact manifold, the operator $\mathcal{O}_{\mbox{\tiny $z$}}$
has a zero vev and the supersymmetries are not broken. This is the only possible configurations for 
theories with no defect ({\it i.e. D$p$/D$(p+4)$ systems}).As pointed out in Section \ref{DefTh}, regularity 
condition at the horizon forces the embedding function to have a trivial profile. We therefore need to
consider the supersymmetric case. A similar analysis was carried out in \cite{Karch:2009eb}. We will
begin with following \cite{Karch:2008fa, Karch:2009eb} by computing thermodynamical quantities such as
the entropy density and the specific heat.

Let us turn on a non-trivial profile for the world-volume gauge field using the ansatz \eqref{FmnAns}
so that the dual gauge theory has a chemical potential. The DBI action and the equation of motion for the
gauge potential $f\left(r\right)$ are
\begin{equation}\eqlabel{DBIfml}
 \begin{split}
 &S_{\mbox{\tiny D$(p+4-2k)$}}\:=\:
  -M\,T_{\mbox{\tiny D$(p+4-2k)$}}\int dt\,d^{\mbox{\tiny $p-k$}}x\,dr\,d^{\mbox{\tiny $3-k$}}\varphi\:
   e^{-\phi}\sqrt{-\mbox{det}\left\{g_{\mbox{\tiny $AB$}}+F_{\mbox{\tiny $AB$}}\right\}}\:=\\
 &\phantom{S_{\mbox{\tiny D$(p+4-2k)$}}\:}=\:
  -M\,T_{\mbox{\tiny D$(p+4-2k)$}}\,g_{\mbox{\tiny $s$}}^{-1}\,\mbox{Vol}\left\{S^{\mbox{\tiny $3-k$}}\right\}
   \int  dt\,d^{\mbox{\tiny $p-k$}}x\,dr\:r^{\mbox{\tiny $3-k$}}\sqrt{1-\left[f'\left(r\right)\right]^2}\\
 &f'\left(r\right)\:=\:\frac{c_{\mbox{\tiny $f$}}}{\sqrt{c_{\mbox{\tiny $f$}}^2+r^{2(3-k)}}}.
 \end{split}
\end{equation}
The contribution to the grand-potential from the fundamental degrees of freedom is given, up to a sign,
by the renormalized on-shell action
\begin{equation}\eqlabel{GPhql}
  \Omega_{\mbox{\tiny fun}}\:=\:-\lim_{\Lambda\rightarrow\infty}
   \left[
    \left.S^{\mbox{\tiny $\Lambda$}}\right|_{\mbox{\tiny on-shell}}+
    \left.S^{\mbox{\tiny $\Lambda$}}\right|_{\mbox{\tiny ct}}
   \right]
\end{equation}
(the holographic renormalization of probe branes in non-conformal backgrounds was extensively studied in
\cite{Benincasa:2009ze}). We can directly compute $\Omega_{\mbox{\tiny fun}}$ in the low-temperature limit
({\it i.e.} $r_{\mbox{\tiny h}}\rightarrow0$), which is the regime we are interested in
\begin{equation}\eqlabel{GPhqlL}
 \begin{split}
  \Omega_{\mbox{\tiny fun}}\:&=\:\Omega_{\mbox{\tiny fun}}^{\mbox{\tiny ($T=0$)}}-
    M\,T_{\mbox{\tiny D$(p+4-2k)$}}\hat{\mathcal{N}}_{\mbox{\tiny $p-k$}}
     \int_{0}^{r_{\mbox{\tiny h}}}dr\:\frac{r^{2(3-k)}}{\sqrt{r^{2(3-k)}+c_{\mbox{\tiny $f$}}^2}}+
    M\,T_{\mbox{\tiny D$(p+4-2k)$}}\hat{\mathcal{N}}_{\mbox{\tiny $p-k$}}
     \frac{r_{\mbox{\tiny h}}^{4-k}}{2(4-k)}\:=\:\\
    &\overset{\mbox{\tiny $T\rightarrow0$}}{=}\:\Omega_{\mbox{\tiny fun}}^{\mbox{\tiny ($T=0$)}}-
      \frac{M\,T_{\mbox{\tiny D$(p+4-2k)$}}\hat{\mathcal{N}}_{\mbox{\tiny $p-k$}}}{2(3-k)+1}
      \frac{r_{\mbox{\tiny h}}^{2(3-k)+1}}{c_{\mbox{\tiny $f$}}}
      \left[1+
      \mathcal{O}\left(\frac{r_{\mbox{\tiny h}}^{2(3-k)}}{c_{\mbox{\tiny $f$}}^2}\right)\right]+\\
    &\phantom{\overset{\mbox{\tiny $T\rightarrow0$}}{=}}+\:
      M\,T_{\mbox{\tiny D$(p+4-2k)$}}\hat{\mathcal{N}}_{\mbox{\tiny $p-k$}}
       \frac{r_{\mbox{\tiny h}}^{4-k}}{2(4-k)},
 \end{split}
\end{equation}
where $\Omega_{\mbox{\tiny fun}}^{\mbox{\tiny ($T=0$)}}$ is the grand-potential at zero temperature,
which is explicitly provided in Section \ref{cpaxis} (setting $c_{\mbox{\tiny $y$}}=0$) and we rewrite
here for future convenience:
\begin{equation}\eqlabel{GP0}
 \Omega_{\mbox{\tiny fun}}^{\mbox{\tiny ($T=0$)}}\:=\:
  -\frac{M\,T_{\mbox{\tiny D$(p+4-2k)$}}\hat{\mathcal{N}}_{\mbox{\tiny $p-k$}}}{(4-k)\mathtt{a}^{3-k}}
   \left(\mu^{\mbox{\tiny ($T=0$)}}\right)^{4-k}.
\end{equation}
Furthermore, the constant $\hat{\mathcal{N}}_{\mbox{\tiny $p-k$}}$ is a redefinition of 
$\mathcal{N}_{\mbox{\tiny $k$}}$ by including the volume of the defect: 
$\hat{\mathcal{N}}_{\mbox{\tiny $p-k$}}\,\overset{\mbox{\tiny def}}{=}\,
 \mathcal{N}_{\mbox{\tiny $k$}}V_{\mbox{\tiny $p-k$}}\,=\,
 \mathcal{N}_{\mbox{\tiny $k$}}\int d^{\mbox{\tiny $p-k$}}x$. Notice that full grand-potential for
the system is given by the contribution \eqref{GPhqlL} from the fundamental degrees and 
the one from the adjoint ones: $\Omega_{\mbox{\tiny adj}}\:=\:
 -(5-p)N^2\varpi\lambda^{\frac{p-3}{5-p}}T^{2\frac{7-p}{5-p}}$\footnote{The quantity $\varpi$ is just a 
constant dependent only on $p$:
$$
 \varpi\:=\:\left[\frac{2^{29-5p}\pi^{13-3p}}{(7-p)^{3(7-p)}}\Gamma\left(\frac{7-p}{2}\right)
  \right]^{\frac{2}{5-p}}
$$}. 
The latter contribution, as well as the last term in \eqref{GPhqlL} are 
independent of the baryon density. Therefore, in order to study the features of the ``quantum liquids''
we need to focus just on the first two terms of 
\eqref{GPhqlL} which will be indicated as $\hat{\Omega}_{\mbox{\tiny fun}}$.

Similarly, we can compute the chemical potential in the low-temperature limit by integrating the
equation of motion \eqref{DBIfml} for the gauge potential $f(r)$:
\begin{equation}\eqlabel{CPhqlL}
 \begin{split}
  \mu\:&=\mu_{\mbox{\tiny $0$}}-\int_{0}^{\mbox{\tiny $r_{\mbox{\tiny h}}$}}
        \frac{c_{\mbox{\tiny $f$}}}{\sqrt{r^{2(3-k)}+c_{\mbox{\tiny $f$}}^2}}\:=\\
  &\overset{\mbox{\tiny $T\rightarrow0$}}{=}\:
    \mu_{\mbox{\tiny $0$}}-r_{\mbox{\tiny h}}
    \left[1-\frac{1}{2\left[2(3-k)+1\right]}\frac{r_{\mbox{\tiny h}}^{2(3-k)}}{c_{\mbox{\tiny $f$}}^2}+
      \mathcal{O}\left(\frac{r_{\mbox{\tiny h}}^{4(3-k)}}{c_{\mbox{\tiny $f$}}^4}\right)\right],
 \end{split}
\end{equation}
with $\mu_{\mbox{\tiny $0$}}\:=\:\mathtt{a}\,c_{\mbox{\tiny $f$}}^{\frac{1}{3-k}}$ being the chemical
potential at zero temperature. Notice that the parameter expansion in \eqref{GPhqlL} and \eqref{CPhqlL} is the
dimensionless ratio $r_{\mbox{\tiny h}}/c_{\mbox{\tiny $f$}}^{\frac{1}{3-k}}$ or, equivalently in terms of the
temperature and the chemical potential at zero temperature,  $T^{\frac{2}{5-p}}/\mu_{\mbox{\tiny $0$}}$.

The entropy density can be now computed as a function of $T$ and $c_{\mbox{\tiny $f$}}$.
\begin{equation}\eqlabel{ShqlL}
 \begin{split}
  s\left(T,\,c_{\mbox{\tiny $f$}}\right)&\:=\:
  -\frac{1}{V_{\mbox{\tiny $p-k$}}}
   \left.\frac{\partial\hat{\Omega}_{\mbox{\tiny fun}}}{\partial T}\right|_{\mbox{\tiny $\mu$}}\:=\:
   -\frac{1}{V_{\mbox{\tiny $p-k$}}}
   \left[
   \left.\frac{\partial\hat{\Omega}_{\mbox{\tiny fun}}}{\partial T}\right|_{
    \mbox{\tiny $c_{\mbox{\tiny $f$}}$}}-
    \left.\frac{\partial\hat{\Omega}_{\mbox{\tiny fun}}}{\partial 
     c_{\mbox{\tiny $f$}}}\right|_{\mbox{\tiny $T$}}
   \frac{\left.\frac{\partial\mu}{\partial T}\right|_{\mbox{\tiny $c_{\mbox{\tiny $f$}}$}}}{
   \left.\frac{\partial\mu}{\partial c_{\mbox{\tiny $f$}}}\right|_{\mbox{\tiny $T$}}}\right]\:=
   \\
   &=\:M T_{\mbox{\tiny D$(p+4-2k)$}}\mathcal{N}_{\mbox{\tiny $k$}}
   \frac{2}{5-p}\left(\frac{4\pi}{7-p}\right)^{\frac{2}{5-p}}
   c_{\mbox{\tiny $f$}}T^{\frac{p-3}{5-p}}\times\\
   &\phantom{=\:}\times
   \left[
    1+\left(\frac{4\pi}{7-p}\right)^{4\frac{3-k}{5-p}}
       \frac{T^{4\frac{3-k}{5-p}}}{2c_{\mbox{\tiny $f$}}^2}-
      \left(\frac{4\pi}{7-p}\right)^{8\frac{3-k}{5-p}}
       \frac{T^{8\frac{3-k}{5-p}}}{8c_{\mbox{\tiny $f$}}^4}+
    \mathcal{O}\left(\frac{T^{12\frac{3-k}{5-p}}}{c_{\mbox{\tiny $f$}}^6}\right)
   \right].
 \end{split}
\end{equation}
The leading term in the entropy density in $s\left(T,\,c_{\mbox{\tiny $f$}}\right)$ scales with the 
temperature as $s\,\sim\,T^{\mbox{\tiny $\frac{p-3}{5-p}$}}$, which is independent of the codimensionality of
the defect. Notice that actually this term is of order one in the conformal case $p\,=\,3$, while for
$p\,<\,3$ it decreases with the temperature as $s\,\sim\,T^{-\frac{3-p}{5-p}}$ and for $p\,=\,4$ it increases 
linearly as $s\,\sim\,T$. This would imply that only in the conformal case, the entropy at zero temperature
is non-zero. Not only. In the limit $T\,\rightarrow0\,$ the entropy density seems to blow up for $p\,<\,3$.
We will comment later on this.

From the entropy density \eqref{ShqlL}, it is straightforward to compute the specific heat 
$c_{\mbox{\tiny V}}$ at constant volume and density (in the limit of low temperature).
\begin{equation}\eqlabel{CVhqlL}
 \begin{split}
  c_{\mbox{\tiny V}}\:&=\:T\left.\frac{\partial s}{\partial T}\right|_{\mbox{\tiny $c_{\mbox{\tiny $f$}}$}}
   \:=\\
  &=\:M T_{\mbox{\tiny D$(p+4-2k)$}}\mathcal{N}_{\mbox{\tiny $k$}}
   \frac{2}{5-p}\left(\frac{4\pi}{7-p}\right)^{\frac{2}{5-p}}
   \frac{T^{\frac{p-3}{5-p}}}{c_{\mbox{\tiny $f$}}}
   \left[
    \frac{p-3}{5-p}+
    \frac{9+p-4k}{2(5-p)}\left(\frac{4\pi}{7-p}\right)^{4\frac{3-k}{5-p}}
     \frac{T^{4\frac{3-k}{5-p}}}{c_{\mbox{\tiny $f$}}^2}-\right.\\
   &\phantom{=\:\frac{p-3}{5-p}}-
    \left.
    \frac{21+p-8k}{8(5-p)}\left(\frac{4\pi}{7-p}\right)^{8\frac{3-k}{5-p}}
     \frac{T^{8\frac{3-k}{5-p}}}{c_{\mbox{\tiny $f$}}^4}+
    \mathcal{O}\left(\frac{T^{12\frac{3-k}{5-p}}}{c_{\mbox{\tiny $f$}}^6}\right)
   \right].
 \end{split}
\end{equation}
For $p\,=\,3$ the first term in \eqref{CVhqlL} vanishes	so that the leading term for the specific heat
$c_{\mbox{\tiny V}}$ scales with the temperature as $T^{2(3-k)}$ (see \cite{Karch:2008fa}). 
For $p\,=\,4$, $c_{\mbox{\tiny $V$}}\,\sim\,T$ independently of the dimension of the defect in which
the fundamental massless degrees of freedom propagates. This is actually the same behaviour of Fermi
liquids. It is interesting to notice that the leading term for the specific heat 
$c_{\mbox{\tiny V}}$ becomes negative for $p\,<\,3$. This would be indeed a signature of a thermodynamical
instability in the canonical ensemble. However, our analysis is carried out in the grand canonical
ensemble where the thermodynamical stability of the system is guaranteed if and only if the Hessian of
(minus) the grand potential is positive definite
\begin{equation}\eqlabel{Hessian}
 \mathcal{H}\left(\hat{\Omega}_{\mbox{\tiny fun}}\right)\:\equiv\:
 \frac{\partial^2\left(-\hat{\Omega}_{\mbox{\tiny fun}}\right)}{\partial\mathcal{X}_{\mbox{\tiny $i$}}
  \partial\mathcal{X}_{\mbox{\tiny $j$}}}\:\equiv\:
  \begin{pmatrix}
   \Omega_{\mbox{\tiny $TT$}} & \Omega_{\mbox{\tiny $T\mu$}} \\
   \Omega_{\mbox{\tiny $T\mu$}} & \Omega_{\mbox{\tiny $\mu\mu$}}
  \end{pmatrix},
  \qquad 
  \mathcal{X}\:=\:\left\{T,\,\mu\right\},
\end{equation}
where the elements 
$\left\{\Omega_{\mbox{\tiny $TT$}}, \Omega_{\mbox{\tiny $T\mu$}}, \Omega_{\mbox{\tiny $T\mu$}}, 
\Omega_{\mbox{\tiny $\mu\mu$}}\right\}$ of the Hessian \eqref{Hessian} are explicitly given by
\begin{equation}\eqlabel{Hessian2}
 \begin{split}
 &\Omega_{\mbox{\tiny $TT$}}\:\equiv\:\left.\frac{\partial^2\left(-\Omega\right)}{\partial T^2}
   \right|_{\mbox{\tiny $\mu$}}\:=\:
  \frac{2}{(5-p)}\tilde{\mathcal{N}}\left(\frac{4\pi}{7-p}\right)^{\frac{2}{5-p}}
  c_{\mbox{\tiny $f$}}T^{-2\frac{4-p}{5-p}}\times\\
 &\phantom{\Omega_{\mbox{\tiny $TT$}}\:\equiv\:}\times 
  \left[
   \frac{p-3}{5-p}+2\frac{3-k}{5-p}\left(\frac{4\pi}{7-p}\right)^{4\frac{3-k}{5-p}}
    \frac{T^{\frac{2}{5-p}}}{\mu_{\mbox{\tiny $0$}}}+
   \frac{9+p-4k}{2(5-p)}\left(\frac{4\pi}{7-p}\right)^{8\frac{3-k}{5-p}}
    \frac{T^{2\frac{3-k}{5-p}}}{c_{\mbox{\tiny $f$}}^2}+\right.\\
 &\phantom{\Omega_{\mbox{\tiny $TT$}}\:\equiv\:\frac{p-3}{5-p}}\left.+\,
    \mathcal{O}\left(\frac{T^{2\frac{7-2k}{5-p}}}{\mu_{\mbox{\tiny $0$}}c_{\mbox{\tiny $f$}}^2}\right)
  \right]\\
 &\Omega_{\mbox{\tiny $T\mu$}}\:\equiv\:\frac{\partial^2\left(-\Omega\right)}{\partial T\partial\mu}\:=\:
  2\frac{3-k}{5-p}\tilde{\mathcal{N}}\left(\frac{4\pi}{7-p}\right)^{\frac{2}{5-p}}
  \frac{c_{\mbox{\tiny $f$}}}{\mu_{\mbox{\tiny $0$}}}T^{\frac{p-3}{5-p}}\times\\
 &\phantom{\Omega_{\mbox{\tiny $T\mu$}}\:\equiv\:}\times
  \left[
   1-\left(\frac{4\pi}{7-p}\right)^{4\frac{3-k}{5-p}}\frac{T^{4\frac{3-k}{7-p}}}{2c_{\mbox{\tiny $f$}}^2}+
     \frac{3-k}{2(3-k)+1}\left(
     \frac{4\pi}{7-p}\right)^{2\frac{7-2k}{5-p}}
    \frac{T^{2\frac{7-2k}{5-p}}}{\mu_{\mbox{\tiny $0$}}c_{\mbox{\tiny $f$}}^2}+\right.\\
  &\phantom{\Omega_{\mbox{\tiny $T\mu$}}\:\equiv\:(p-3)}
     \left.+\frac{3}{8}
      \left(\frac{4\pi}{7-p}\right)^{8\frac{3-k}{5-p}}\frac{T^{8\frac{3-k}{7-p}}}{c_{\mbox{\tiny $f$}}^2}+
    \mathcal{O}\left(\frac{T^{2\frac{13-2k}{5-p}}}{\mu_{\mbox{\tiny $0$}}c_{\mbox{\tiny $f$}}^4}\right)
  \right]\\
 &\Omega_{\mbox{\tiny $\mu\mu$}}\:\equiv\:
  \left.\frac{\partial^2\left(-\Omega\right)}{\partial\mu^2}\right|_{\mbox{\tiny $T$}}\:=\:
   \tilde{\mathcal{N}}(3-k)\frac{c_{\mbox{\tiny $f$}}}{\mu_{\mbox{\tiny $0$}}}\times\\
 &\phantom{\Omega_{\mbox{\tiny $\mu^2$}}\:\equiv\:}
  \times
  \left[
   1+\frac{3-k}{2(3-k)+1}\left(\frac{4\pi}{7-p}\right)^{2\frac{7-2k}{5-p}}
    \frac{T^{2\frac{7-2k}{5-p}}}{\mu_{\mbox{\tiny $0$}}c_{\mbox{\tiny $f$}}^2}-\right.\\
 &\phantom{\Omega_{\mbox{\tiny $\mu^2$}}\:\equiv\:p-3}
    \left.-\frac{3}{2\left[4(3-k)+1\right]}\left(\frac{4\pi}{7-p}\right)^{2\frac{13-2k}{5-p}}
    \frac{T^{2\frac{13-2k}{5-p}}}{\mu_{\mbox{\tiny $0$}}c_{\mbox{\tiny $f$}}^4}+\ldots
  \right],
 \end{split}
\end{equation}
where $\tilde{\mathcal{N}}\:=\:M T_{\mbox{\tiny D$(p+4-2k)$}}\hat{\mathcal{N}}_{\mbox{\tiny $p-k$}}$.
In order to have the stability of the system guaranteed, the Hessian 
$\mathcal{H}\left(\hat{\Omega}_{\mbox{\tiny fun}}\right)$ in \eqref{Hessian} needs to be (semi)-definite 
positive, which means that the eigenvalues need to be positive (or at most zero). This is also equivalent
to requiring the non-negativity of all the principal minors of 
$\mathcal{H}\left(\hat{\Omega}_{\mbox{\tiny fun}}\right)$. Being a $2\times2$ matrix, there are just two
principal minors: $\Omega_{\mbox{\tiny $TT$}}$ and the determinant $\left|\mathcal{H}\right|$ of the
Hessian. From the explicit expression in \eqref{Hessian2}, the smallest minor $\Omega_{\mbox{\tiny $TT$}}$
becomes negative for $p\,<\,3$. As far as the determinant 
$\left|\mathcal{H}\right|$ is concerned, it has the following form:
\begin{equation}\eqlabel{detH}
 \begin{split}
 \left|\mathcal{H}\right|\:&=\:
  2\frac{3-k}{5-p}\tilde{\mathcal{N}}^2\left(\frac{4\pi}{7-p}\right)^{\frac{2}{5-p}}
   \frac{c_{\mbox{\tiny $f$}}^2}{\mu_{\mbox{\tiny $0$}}}T^{-2\frac{4-p}{5-p}}
  \left[
   \frac{p-3}{5-p}+\frac{9+p-4k}{2(5-p)}\frac{T^{4\frac{3-k}{5-p}}}{c_{\mbox{\tiny $f$}}^2}+\ldots
  \right]
 \end{split}
\end{equation}
As for the minor $\Omega_{\mbox{\tiny TT}}$, the determinant \eqref{detH} has the leading term in the
low temperature expansion which becomes negative for $p\,<\,3$. Therefore, the Hessian turns out to be
negative definite for $p\,<\,3$. In the conformal case $p\,=\,3$, the first term in \eqref{detH} vanishes and
the new dominant contribution is always positive.
Finally, for $p\,=\,4$ all the terms in \eqref{detH} are positive. The Hessian is therefore positive definite
for $p\,\ge\,3$. The positive-definiteness of the Hessian for the cases $p\,=\,3,\,4$ insures stability. 
Moreover, the entropy density at zero temperature is either finite ($p=3$) or zero $p=4$, in agreement with 
the third law of thermodynamics. The specific heat turns out to behave as $\sim\,T^{2(3-k)}$ ($p=3$) or
$\sim\,T$ ($p=4$).

A comment is now in order. The statement of the negative-definiteness of the Hessian is a local statement.
This means that, in principle, the systems could tend to a stable configuration. Moreover, the brane 
configurations analysed in this section show a very peculiar behaviour for
entropy density \eqref{ShqlL} and specific heat \eqref{CVhqlL} at low temperature for $p\,<\,3$. 
For convenience, let us rewrite it below:
\begin{equation}\eqlabel{SCVhqlL}
 s\left(T,\,c_{\mbox{\tiny $f$}}\right)\:=\:\tilde{\mathcal{N}}\frac{2}{5-p}
  \left(\frac{4\pi}{7-p}\right)^{\frac{2}{5-p}}c_{\mbox{\tiny $f$}}\,T^{-\frac{3-p}{5-p}}+\ldots,
 \qquad
 c_{\mbox{\tiny $V$}}\:=\:\tilde{\mathcal{N}}2\frac{p-3}{(5-p)^2}\left(\frac{4\pi}{7-p}\right)^{\frac{2}{5-p}}
  \frac{T^{-\frac{3-p}{5-p}}}{c_{\mbox{\tiny $f$}}}+\ldots
\end{equation}
The behaviour \eqref{SCVhqlL} implies that both entropy density and specific heat increase as the temperature
approaches to zero. This seems to violate the third law of thermodynamics according to which the density
entropy reaches a minimum value as the temperature approaches to zero. As a consequence, the specific heat
should vanish in the same limit. This, together with the negative-definiteness of the Hessian (which, at the
end of the day, is a consequence of the behaviour \eqref{SCVhqlL}), implies that the configuration analysed
here are not thermodynamically stable for $p\,<,3$.  However, we need to recall
that zero temperature backreacted solutions do exist for $p\,<\,3$ which are well-behaved, like the D$2$/D$6$ 
solution found in \cite{Cherkis:2002ir}. It is therefore natural to ask how the appearance of this 
low-temperature instability connects with the existence of well-behaved backreacted solutions. One
explanation to this question can be provided by the fact that these systems can become stable once
one goes beyond the probe approximation, {\it i.e.} it is the the backreaction which stabilises them.
The only tunable parameter available is the number $M$ of probe branes. One can thus think to increase the
number of D$(p+4-2k)$-branes until the Hessian becomes positive definite and thus the systems stabilise. This 
would imply a modification of the potential curve and therefore the effect of increasing $M$ can't be seen 
from our computation, given that the potential curves of the probe case and of the backreacted one are 
different. In other words, tuning $M$ one gets a family of potentials until stability is reached
at the backreaction, when the probe approximation breaks down.

One might argue that we are just computing thermodynamical quantities which refer just to the fluid 
({\it i.e. } to the fundamental degrees of freedom), and that our stability analysis does not take into
account all the degrees of freedom. This can be done by starting from the full grand potential, which
contains both a contribution from the adjoint degrees of freedom and from the fundamental ones (including
the density independent term in \eqref{GPhqlL}). The contributions from $\Omega_{\mbox{\tiny adj}}$ scales 
with the temperature as $T^{2\frac{7-p}{5-p}}$, while the last term in \eqref{GPhqlL} as 
$T^{2\frac{4-k}{5-p}}$. It easy to see that such a contribution will not affect the leading behaviour at low 
temperature of any of the relevant thermodynamical quantities, such as the density entropy, the specific 
heat and the Hessian of the grand potential. Thus the instability we are observing is not indeed due to
not keeping into account the whole grand potential.

In principle, looking at the eigenvalues of the Hessian, one might think that the system can be driven
to a high temperature stable point. In this case, the contribution from the adjoint degrees of freedom becomes
more and more important as the temperature increases. 
Simultaneously the probe branes get heat up and can acquire enough stress-energy to
eventually backreact. The probe approximation may break down anyway. It seems reasonable to think
that the backreaction can stabilise the system, and the negative-definite Hessian, together with an
apparent violation of the third law of thermodynamics, is a signature of the breaking down
of the probe approximation.

We would like to stress that a deeper analysis of such an instability is indeed needed, since our arguments
do not provide a robust proof of the nature of such an instability as the breaking-down of the 
probe approximation.

\section{Conclusion}\label{Concl}

In this paper, we investigate the phase structure of D$p$/D$(p+4-2k)$-systems, where the D$(p+4-2k)$
branes are considered in the probe approximation. These systems are BPS and introduce a $(p+1-k)$-dimensional
defect in the $(p+1)$-dimensional $U(N)$ gauge theory. We consider both the probe brane configurations
which introduce a massive fundamental hypermultiplet and the ones which introduce massless excitations in the 
$(p+1-k)$-dimensional defect. 

The probe D$(p+4-2k)$-branes wrap a $\mathcal{M}_{\mbox{\tiny $p+2-k$}}\times S^{3-k}$ subspace of the
background geometry $\mathcal{M}_{\mbox{\tiny $p+2$}}\times S^{8-p}$ generated by a stack of D$p$-branes.
The embedding of the probe branes can be controlled by a scalar mode, which is provided by one of the
angular coordinates in the transverse space. The separation between the stack of probe branes and the
background ones is a parameter in the theory related to the quark mass.
In this setting we consider the system to be at finite temperature ({\it i.e.} non-extremal
black hole background) and at finite chemical potential, by turning on a non-trivial profile for the
gauge field on the probe branes world-volume. On the chemical potential axis ($T\,=\,0$), the system undergoes
a second order phase transition for $k\,=\,0,\,1$. In the region $\mu\,<\,m_{\mbox{\tiny $0$}}$ the probe 
branes cannot reach the location of the probe branes, but they have an extreme point where they turn back and
hit the boundary again: the probe branes are in a D$(p+4-2k)$/$\bar{\mbox{D$(p+4-2k)$}}$ configuration.
For $\mu\,>\,m_{\mbox{\tiny $0$}}$ the probe brane can reach the location of the background branes, in
a black-hole crossing phase. The second phase is thermodynamically favoured since the grand-potential is 
negative, while in the brane/anti-brane phase it is positive. The case $k\,=\,2$ is a bit special given its
different physical interpretation: turning on a non-trivial embedding mode in the transverse space does no longer correspond to a mass-deformation, {\it i.e. } the separation between the probe branes and the background
one can no longer be interpreted as a mass, but it provides a non-zero vacuum-expectation-value for its dual 
operator.

In the interior of the $\left(\mu,T\right)$-plane, it is possible to identify a transition-curve
$\mu\,=\,m\left(T\right)$ between Minkowski embeddings and black-hole embeddings. The order of the
phase transition across this curve is strongly tied to the codimensionality $k$ of the defect. In the
case of $k=0$, the result of \cite{Faulkner:2008hm} can be extended to any {\it sensible} 
D$p$/D$(p+4)$-systems: the transition across the curve $\mu\,=\,m\left(T\right)$ is a third order phase
transitions. Third-order phase transitions are not very common in nature. However, there are some 
meaningful examples to keep into consideration. One example is indeed provided by the Gross-Witten model
\cite{Gross:1980he} in which a third-order phase transition in the large-$N$ lattice gauge theory in 
two-dimensions was observed and then extended to four-dimensions. Indeed, the physics of the phase transition
in the Gross-Witten model is very different from the one of the class of theories discussed in this paper.
It is a weak-to-strong coupling phase transition which occurs at fixed 't Hooft coupling 
($\lambda\,=\,\lambda_{\mbox{\tiny $c$}}$) and at large-$N$. Here instead, the system is studied at both 
large-$N$ and infinite 't Hooft coupling and the transition occurs in the strongly-coupled regime 
(for $p\,\ge,0$). The gauge theories dual to these brane construction define a new class of theories with a 
third order phase transition.

For such systems, the transition line $\mu\,=\,m\left(T\right)$ represents a third order phase transition
until it reaches the temperature $T\,=\,T_{\mbox{\tiny c}}$ at which the phase transition becomes of second
order.

In the case $k=1$, the phase transition along $\mu\,=\,m\left(T\right)$ is of second order. This is
a direct consequence of the fact that, for such systems, the small ``quark''-density expansion of the
chemical potential does not show any logarithmic behaviour at first order in ``quark''-density.

Theories with a codimension-$1$ defect can be reduced to effective codimension-$0$ theories by compactifying
the direction of the background on which the probe branes do not extend. An example of such systems 
was studied in \cite{Kruczenski:2003uq, Mateos:2007vn, Matsuura:2007zx}. One can wonder if and how the
reduction of theories with a codimension-$1$ defect to effective theories with no defect affects the 
order of the phase transition along the curve $\mu\:=\:m(T)$. Most likely, it is the case because of the
introduction of a new scale provided by the compactification radius. However, we did not check this
explicitly and we leave it for future work. Indeed, if the compactification has the effect of changing
the order of the phase transition along $\mu\:=\:m(T)$, it would strengthen our statement of the
existence of two universality classes of theories with baryonic chemical potential, which are identified
by the codimensionality of the defect, {\it i.e. }  by the order of the phase transition.

The massless degrees of freedom on the $(p+1-k)$-dimensional defect are studied by fixing the probe branes
to wrap the maximal $(3-k)$-sphere in the transverse space. Fixing also the position of the probe-branes
in the non-compact directions provides a supersymmetric description of the system. In principle one can turn
on a non-trivial profile for $x^{\mbox{\tiny $p$}}=z\left(\rho\right)$ (indeed this description is valid
for $k\neq0$). In this case, the operator $\mathcal{O}_{\mbox{\tiny $z$}}$ would acquire a non-zero
vev and the supersymmetries get broken. We consider the supersymmetric case and focus on the low temperature
properties of such finite density systems.
Following \cite{Karch:2008fa, Karch:2009eb}, we compute the density entropy and the specific heat in such
a regime. Both those quantities turn out to scale with the temperature as
\begin{equation}\eqlabel{scvC}
s\:\sim\:T^{\frac{p-3}{5-p}}, \qquad c_{\mbox{\tiny $V$}}\:\sim\:(p-3)T^{\frac{p-3}{5-p}}.
\end{equation}
It is interesting to notice that the powers in \eqref{scvC} become negative for $p\,<\,3$. Indeed, this
type of behaviour seems to be at least counter-intuitive. For the third law of thermodynamics one would 
expect the specific heat to vanish as the temperature goes to zero and the density entropy to reach zero as
well or, anyway, a finite value. Instead, for the class of systems identified by $p\,<\,3$ both these
quantities appear to blow up in the zero temperature limit. In order to investigate this issue, we explicitly
analyse the stability of these systems. A necessary and sufficient condition for the (local) stability of
these systems is that the Hessian matrix is positive definite. In the case of interest ($p\,<\,3$), we show 
that the Hessian is actually negative definite, which identifies an instability at low temperature explaining
the observed behaviour of entropy and specific heat. However, it arises new questions. First, what is the
nature of such an instability? How can these systems be stabilised? How does this result connects with
the existence of a zero-temperature backreacted solution such as the D$2$/D$6$ of \cite{Cherkis:2002ir}?
We pointed out that the only tunable parameter in the theory is the number $M$ of probe branes. So one can 
think to increase it until the potential changes in such a way to make the Hessian positive definite, at
which point most likely the $M$ D$(p+4-2k)$-branes ($p\,<\,3$) will start backreacting stabilising the system 
and the probe approximation breaks down. Another possibilities is that the analysis of the eigenvalues of the 
Hessian may suggest that the system can be lead to a stable point at high temperature. In this case, 
the contribution from the adjoint degrees of freedom as well as from the density independent term
from the fundamental ones will start to be more and more relevant. Simultaneously, the probe branes
get heat up to the point that they can start to backreact. This would lead again to the breaking-down
of the probe approximation. It is therefore reasonable to think that the backreaction is needed to
stabilise the system, and consider it as best candidate for explaining the appearance of such an instability
at low temperature. Once again, we want to reiterate the idea that a more detailed analysis about this
stability issue is needed, but our observation stresses the necessity of an extensive discussion about
the validity of the probe approximation as well as the need of a deeper understanding of backreacted solutions
and the related physics.

\section*{Acknowledgment}

It is a pleasure to thank Michael Haack and Suresh Nampuri for hospitality at LMU as well as the organisers
of the Workshop on the Fluid Gravity Correspondence held in Munich and $5^{\mbox{\tiny $th$}}$ Aegean Summer 
School in Milos for the stimulating environment. I am very thankful to Suresh Nampuri for discussions and
correspondence. I would like to thank Johanna Erdmenger, Michael Haack and the string theory groups at LMU and
MPI Munich for the possibility to present preliminary results of this work in their local seminar, as well as 
Giuseppe Policastro and the LPTENS for the possibility to present them in the  ``Rencontres 
Th{\'e}oriciennes''. It is also a pleasure to thank for stimulating discussions and interest 
Marcello Dalmonte, Jan de Boer, Pau Figueras, Veronika Hubeny, David Mateos, Mukund Rangamani, Andrea 
Scaramucci. This work is supported by STFC Grant.

\appendix

\section{Coefficients for the small density expansions}\label{App1}

In this section we explicitly write down the numerical coefficients in the small density expansion
of section \ref{Cpsde}, where we emphasised as in \cite{Faulkner:2008hm} that such an expansion is 
subtle and it is necessary to split the radial axis in two regions. In region 1, at zero order
in $c_{\mbox{\tiny $f$}}$ the embedding function is conveniently expanded in a neighbourhood of
$\varrho\,\sim\,0$ and such expansion turns out to involve even powers only of the radial coordinate
\eqref{y0r}. The coefficients of this expansion have been written in \eqref{y0rCoeffs} in terms of
other coefficients $\mathtt{b}_{\mbox{\tiny $i$}}^{\mbox{\tiny $(p,k)$}}$ and 
$\mathtt{c}_{\mbox{\tiny $i$}}^{\mbox{\tiny $(p,k)$}}$ which depend on the spatial dimensions $p$
on the background branes and on the codimensionality $k$ of the defect. Here we list the coefficients
 $\mathtt{b}_{\mbox{\tiny $i$}}^{\mbox{\tiny $(p,k)$}}$ and 
$\mathtt{c}_{\mbox{\tiny $i$}}^{\mbox{\tiny $(p,k)$}}$:
\begin{itemize}
\item Coefficients $\mathtt{b}_{\mbox{\tiny $i$}}^{\mbox{\tiny $(p,k)$}}$ in 
      $y_{\mbox{\tiny $0$}}^{\mbox{\tiny $\left(4\right)$}}$
\begin{equation}\eqlabel{bcoeffs}
 \begin{split}
  &\mathtt{b}_{\mbox{\tiny $0$}}^{\mbox{\tiny $(p,k)$}}\:=\:
   -(4-k)^3 \left(p^2-(12+k)p+9(k+3)\right),\\
  &\mathtt{b}_{\mbox{\tiny $1$}}^{\mbox{\tiny $(p,k)$}}\:=\:
   2(4-k)^2 \left(p^3-2(k+8)p^2+(2k^2+18k+85)p-2(9k^2-2k+109)\right),\\
  &\mathtt{b}_{\mbox{\tiny $2$}}^{\mbox{\tiny $(p,k)$}}\:=\:
   -2(4-k)\left(4p^3-3(k^3-12k^2+52k-52)p^2+\right.\\
  &\phantom{\mathtt{b}_{\mbox{\tiny $2$}}^{\mbox{\tiny $(p,k)$}}\:=\:}
   +\left.(3k^4-2k^3-252k^2+1512-2068)p-(3+k)(25k^3-296k^2+1224k-1564)\right),\\
  &\mathtt{b}_{\mbox{\tiny $3$}}^{\mbox{\tiny $(p,k)$}}\:=\:
   2(4-k)\left((4-k)p^3+2(k^2+k-14)p^2-(2k^3-2k^2+49k-124)p+\right.\\
  &\phantom{\mathtt{b}_{\mbox{\tiny $3$}}^{\mbox{\tiny $(p,k)$}}\:=\:}
   \left.+4(3k^3-19k^2+95k-137)\right),\\
  &\mathtt{b}_{\mbox{\tiny $4$}}^{\mbox{\tiny $(p,k)$}}\:=\:
   -(4-k)^2\left((4-k)p^2+(k^2-4k+24)p-3(12-k)(3+k)\right),\\
  &\mathtt{b}_{\mbox{\tiny $5$}}^{\mbox{\tiny $(p,k)$}}\:=\:
   4(6-k)(4-k)^3.
 \end{split}
\end{equation}
\item Coefficients $\mathtt{c}_{\mbox{\tiny $i$}}^{\mbox{\tiny $(p,k)$}}$ in
      $y_{\mbox{\tiny $0$}}^{\mbox{\tiny $\left(6\right)$}}$
      \begin{equation}\eqlabel{ccoeffs}
       \begin{split}
        &\mathtt{c}_{\mbox{\tiny $0$}}^{\mbox{\tiny $(p,k)$}}\:=\:
          (6-k)(4-k)^5(11-p)(9-p)(3+k-p),\\
        &\mathtt{c}_{\mbox{\tiny $1$}}^{\mbox{\tiny $(p,k)$}}\:=\:
          2(4-k)^6(9-p)
          \left[(16-3k)p^3+(6k^2+17k-264)p^2-\right.\\
        &\phantom{\mathtt{c}_{\mbox{\tiny $1$}}^{\mbox{\tiny $(p,k)$}}\:=\:}
          \left.-(k^3+92k^2-347k-960)p+(15k^3+242k^2-1753k+24)\right],\\
        &\mathtt{c}_{\mbox{\tiny $2$}}^{\mbox{\tiny $(p,k)$}}\:=\:
         4(4-k)^3(3+k-p)
         \left[2(5-k)p^4+4(k^2+29k-256)p^3+\right.\\
        &\phantom{\mathtt{c}_{\mbox{\tiny $2$}}^{\mbox{\tiny $(p,k)$}}\:=\:}
          +(4k^3-156k^2+33k+1258)p^2-(74k^3-1884k^2+7565k-3970)p-\\
        &\phantom{\mathtt{c}_{\mbox{\tiny $2$}}^{\mbox{\tiny $(p,k)$}}\:=\:}
          \left.-(4-k)(340k^2-5848k+9573)\right],\\
        &\mathtt{c}_{\mbox{\tiny $3$}}^{\mbox{\tiny $(p,k)$}}\:=\:
         -2(4-k)^2\left[88p^5+(11k^3-136k^2+208k-2616)p^4-\right.\\
        &\phantom{\mathtt{c}_{\mbox{\tiny $3$}}^{\mbox{\tiny $(p,k)$}}\:=\:}
         -2(11k^4+18k^3-1564k^2+3552k-17176)p^3-\\
        &\phantom{\mathtt{c}_{\mbox{\tiny $3$}}^{\mbox{\tiny $(p,k)$}}\:=\:}
         -(3k^5-714k^4+6150k^3+4896k^2-37152k+209968)p^2+\\
        &\phantom{\mathtt{c}_{\mbox{\tiny $3$}}^{\mbox{\tiny $(p,k)$}}\:=\:}
         +2(2k^5-3147k^4+37726k^3-107556k^2+1458884k+175420)p+\\
        &\phantom{\mathtt{c}_{\mbox{\tiny $3$}}^{\mbox{\tiny $(p,k)$}}\:=\:}
         \left.
          +(203k^5+14898k^4-226797k^3+941720k^2-1807856k+755560)
         \right],\\
        &\mathtt{c}_{\mbox{\tiny $4$}}^{\mbox{\tiny $(p,k)$}}\:=\:
         2(3+k-p)\left[8(2k^4-35k^3+228k^2-648k+696)p^4-\right.\\
        &\phantom{\mathtt{c}_{\mbox{\tiny $4$}}^{\mbox{\tiny $(p,k)$}}\:=\:}
          -8(4k^5-30k^4-299k^3+3940k^2-142898k+17504)p^3+\\
        &\phantom{\mathtt{c}_{\mbox{\tiny $4$}}^{\mbox{\tiny $(p,k)$}}\:=\:}
          +(35k^6-198k^5-1664k^4-3224k^3+197280k^2-924160k+1288320)p^2-\\
        &\phantom{\mathtt{c}_{\mbox{\tiny $4$}}^{\mbox{\tiny $(p,k)$}}\:=\:}
          -4(165k^6-3024k^5+21040k^4-89722k^3+372024k^2-1148640k+1484736)p+\\
        &\phantom{\mathtt{c}_{\mbox{\tiny $4$}}^{\mbox{\tiny $(p,k)$}}\:=\:}
          \left.+(3069k^6-68754k^5+605952k^4-2847248k^3+8357440k^2-15749056k+14671808)\right],\\
        &\mathtt{c}_{\mbox{\tiny $5$}}^{\mbox{\tiny $(p,k)$}}\:=\:
          -2(4-k)\left[8(4-k)p^5-(55k^4-964k^3+5936k^2-17880k+25120)p^4+\right.\\
        &\phantom{\mathtt{c}_{\mbox{\tiny $5$}}^{\mbox{\tiny $(p,k)$}}\:=\:}
          +2(55k^5-424k^4-3884k^3+48176k^2-173528k+237536)p^3-\\
        &\phantom{\mathtt{c}_{\mbox{\tiny $5$}}^{\mbox{\tiny $(p,k)$}}\:=\:}
          -(73k^6+422k^5-15982k^4+42672k^3+415744k^2-2346864k+3647040)p^2+\\
        &\phantom{\mathtt{c}_{\mbox{\tiny $5$}}^{\mbox{\tiny $(p,k)$}}\:=\:}
          +2(560k^6-7005k^5+13600k^4+40732k^3+483088k^2-3681404k+6445680)p-\\
        &\phantom{\mathtt{c}_{\mbox{\tiny $5$}}^{\mbox{\tiny $(p,k)$}}\:=\:}
          \left.-(4187k^6-80082k^5+551079k^4-1930700k^3+5556816k^2-15048120k+20930976)\right]\\
        &\mathtt{c}_{\mbox{\tiny $6$}}^{\mbox{\tiny $(p,k)$}}\:=\:
         4(4-k)^2(3+k-p)\left[2(5-k)(4-k)p^4+(4-k)(4k^2+11k-306)p^3+\right.\\
        &\phantom{\mathtt{c}_{\mbox{\tiny $6$}}^{\mbox{\tiny $(p,k)$}}\:=\:}
         +(32k^4-496k^3+2921k^2-10622k+20264)p^2-\\
        &\phantom{\mathtt{c}_{\mbox{\tiny $6$}}^{\mbox{\tiny $(p,k)$}}\:=\:}
         -(398k^4-7196k^3+4654k^2-149796k+190264)p+\\
        &\phantom{\mathtt{c}_{\mbox{\tiny $6$}}^{\mbox{\tiny $(p,k)$}}\:=\:}
         \left.+(1100k^4-22516k^3+162269k^2-504752k+629168)\right],\\
        &\mathtt{c}_{\mbox{\tiny $7$}}^{\mbox{\tiny $(p,k)$}}\:=\:
         2(4-k)^3\left[3(4-k)(16-3k)p^4+2(4-k)(9k^2+16k-516)p^3+\right.\\
        &\phantom{\mathtt{c}_{\mbox{\tiny $7$}}^{\mbox{\tiny $(p,k)$}}\:=\:}
         +(27k^4-206k^3+450k^2-8200k+33792)p^2-\\
        &\phantom{\mathtt{c}_{\mbox{\tiny $7$}}^{\mbox{\tiny $(p,k)$}}\:=\:}
         -2(136k^4-1896k^3+9704k^2-32996k+79184)p+\\
        &\phantom{\mathtt{c}_{\mbox{\tiny $7$}}^{\mbox{\tiny $(p,k)$}}\:=\:}
         \left.+(405k^4-8014k^3+56877k^2-182228k+322176)\right],
       \end{split}
      \end{equation}
      \begin{equation*}
       \begin{split}
        &\mathtt{c}_{\mbox{\tiny $8$}}^{\mbox{\tiny $(p,k)$}}\:=\:
         (4-k)^2(3+k-p)\left[9(6-k)(4-k)p^2-4(4-k)(154-3k)p+\right.\\
        &\phantom{\mathtt{c}_{\mbox{\tiny $8$}}^{\mbox{\tiny $(p,k)$}}\:=\:}
         \left.+(131k^2-782k+2952)\right],\\
        &\mathtt{c}_{\mbox{\tiny $9$}}^{\mbox{\tiny $(p,k)$}}\:=\:
          8(4-k)^5(11k^2-24k-32).
       \end{split}
      \end{equation*}
\end{itemize}

\bibliographystyle{utphys}
\bibliography{gaugegravityrefs}	

\end{document}